\documentclass{aastex}

\usepackage{longtable}
\usepackage{amssymb}
\usepackage{graphicx}
\usepackage[usenames]{color}
\usepackage{url}
\usepackage{amsmath}
\usepackage{arydshln}

\def \aap {A\&A}

\def \apj {ApJ}
\def \apjl {ApJL}
\def \apjs {ApJS}

\def \mnras {MNRAS}

\def\red#1{\textcolor{red}{ #1}}  % flag for editing
\def\blue#1{\textcolor{blue}{ #1}}  % flag for editing
\shortauthors{Torres \& Lin}
\shorttitle{Superefficiency 
in pulsar wind nebulae }

\begin{document}

\title{Discovery and characterization of superefficiency 
in pulsar wind nebulae}

\author{Diego F. Torres\altaffilmark{1,2,3} \& Tingting Lin\altaffilmark{1,3}}
\altaffiltext{1}{Institute of Space Sciences (ICE, CSIC), Campus UAB, Carrer de Magrans s/n, 08193 Barcelona, Spain}
\altaffiltext{2}{Instituci\'o Catalana de Recerca i Estudis Avan\c{c}ats (ICREA) Barcelona, Spain}
\altaffiltext{3}{Institut d'Estudis Espacials de Catalunya (IEEC), 08034 Barcelona, Spain}

%\date{}
%\maketitle

\begin{abstract}
We numerically study the radiative properties of the reverberation phase of pulsar wind nebulae.
Reverberation brings a significant evolution in a short period of time. 
We show that even the Crab nebula, associated to the more energetic pulsar of the sample we consider, 
has a period in its future time evolution where the X-ray luminosity will exceed the spin-down power at the time. 
In fact, all nebulae in our sample are expected to have a period of radio, X-ray, and GeV superefficiency, and most will also have a period of TeV superefficiency.
We analyze and characterize these superefficient phases.

\end{abstract}


%%%%%%%%%%%%%%%%%%%%%%%%%%%%%%%%%%%%%%%%%%%%%%%%%%
\section{Introduction}
%%%%%%%%%%%%%%%%%%%%%%%%%%%%%%%%%%%%%%%%%%%%%%%%%%

Recently, \cite{Younes2016} reported the discovery of a nebula surrounding the magnetar
Swift J1834.9-0846.
The fact that this system has the highest efficiency of all pulsar wind nebulae (PWNe) known was considered to be highly unusual:
$\sim 10\%$ of the mild spin-down power of the pulsar, 
$L_{sd} \sim 10^{34}$ erg s$^{-1}$, is emitted just in soft X-rays.
This promoted interpretations based on a transfer, via a yet unknown mechanism, of 
 magnetic energy into particle acceleration \citep{Granot2017}.
However, 
we demonstrated that the multifrequency data, as well as its size, could be encompassed 
by a normal, rotationally-powered PWN under the condition that it is 
entering in reverberation \citep{Torres2017}.
The latter is a relatively short but important phase in the evolution of all PWNe, produced when the reverse shock 
created by the supernova explosion travels back toward the pulsar,  compressing the wind bubble,
see, e.g., \cite{Slane2017}, for a review.
This compression heats the PWN, reducing its size, and increasing the magnetic field.
Such evolution leads, as we see below,
to an almost complete burn-off of the electron population.
Despite the obvious importance of this phase, it is not yet usual that radiative models of PWNe consider 
it.
In fact, the effect of reverberation upon the spectral results has been dealt with
only in a few scattered occasions, and with different levels of detail, see, e.g., 
\cite{Gelfand2009,Vorster2013,Bandiera2014,Bucciantini2011,Martin2016,Torres2017}.

Here we aim at studying the radiative properties of the reverberation phase in detail.
For this, we shall study the future reverberation period of well-characterized PWNe.
We shall prove that the 10\% efficiency found for Swift J1834.9-0846 is not a limit at any rate, not even for this very pulsar, 
finding that all PWNe can have periods of superefficiency from radio to gamma-rays.

%%%%%%%%%%%%%%%%%%%%%%%%%%%%%%%%%%%%%%%%%%%%%%%%%%
\section{PWN evolution}
%%%%%%%%%%%%%%%%%%%%%%%%%%%%%%%%%%%%%%%%%%%%%%%%%%

We use the code TIDE 2.3, which has
been described in detail in \cite{Martin2016, Torres2017}.
Here we only add subroutines appropriate to compute efficiencies as a function of time, as described below.
The main components and features of the model, apart that it takes into account the variation of the spin-down power, $L_{sd}$, 
according to a given value of braking index, $n$, are as follows:

\begin{itemize}
\item  The injection function for pairs is assumed as a broken power law, powered by the pulsar. 
The model computes the time-evolution of the distribution subject to 
synchrotron, inverse Compton, and Bremsstrahlung interactions, adiabatic losses or heating, and accounting for escaping particles.
 Expressions for the radiative losses can be found in \cite{Martin2012}.

\item The magnetic field of the nebula is also powered by the rotational power (the instantaneous injection is the fraction of spin-down that goes to power the magnetic field, $\eta$).
The field varies in time as a result of the balance between this power and the adiabatic losses or gains of the field due to the expansion or contraction of the PWN \citep{Torres2013b}.

\item  The size of the PWN is computed according to age, progenitor explosion energy, medium density, velocity, and pressure of the supernova ejecta at the position of the PWN shell. We take into account that the latter profiles change if the PWN shell is surrounded by unshocked ejecta (thus the radius of the PWN is smaller than the radius of the reverse shock of the SNR, $R < R_{rs}$), or by shocked ejecta (where $R_{rs} < R < R_{snr}$, being $R_{snr}$ the radius of the SNR). After reverberation,  when 
the PWN pressure reaches that of the SNR, a  Sedov expansion follows. Details are explicit in \S3 of \cite{Martin2016}.

\end{itemize}

\begin{figure*}
    \centering
\hspace{-0.5cm}
            \includegraphics[width=0.35 \textwidth]{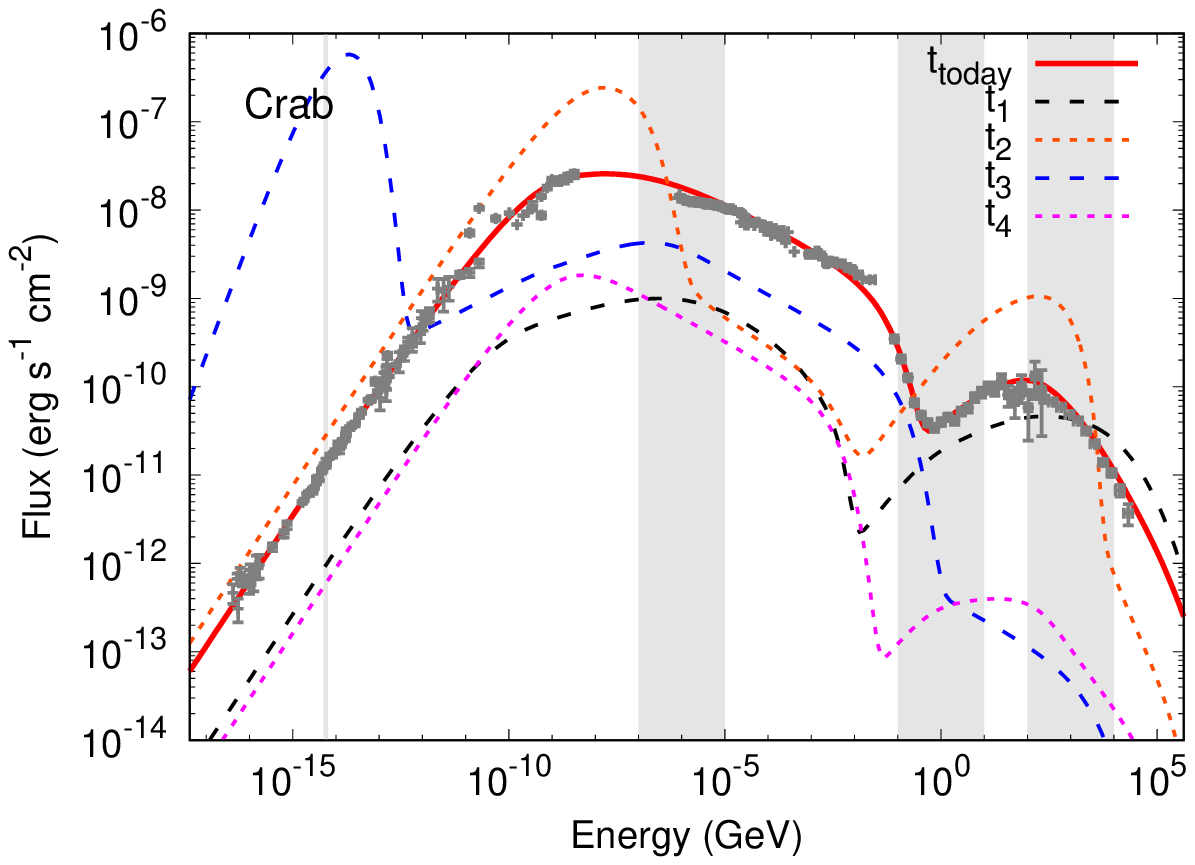}\hspace{-0.4cm}
        \includegraphics[width=0.35 \textwidth]{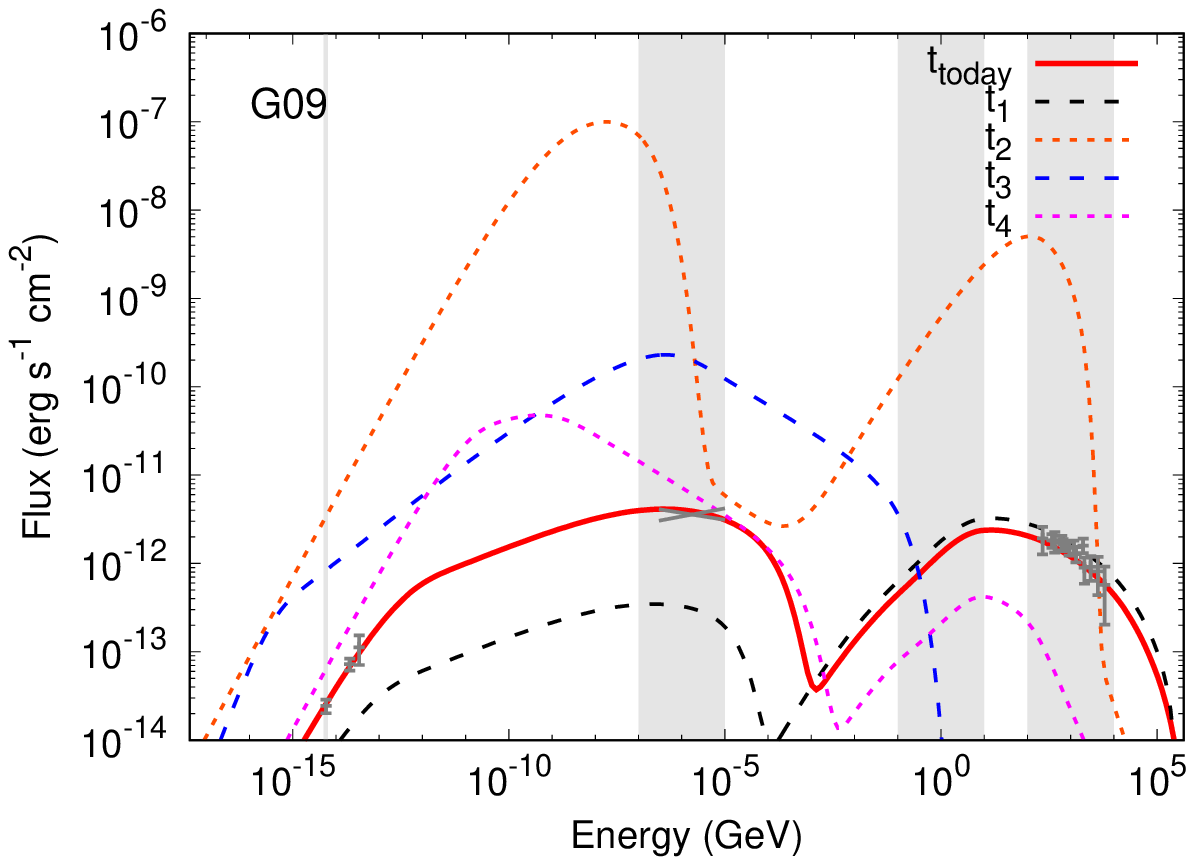} \hspace{-0.5cm}
                \includegraphics[width=0.35 \textwidth]{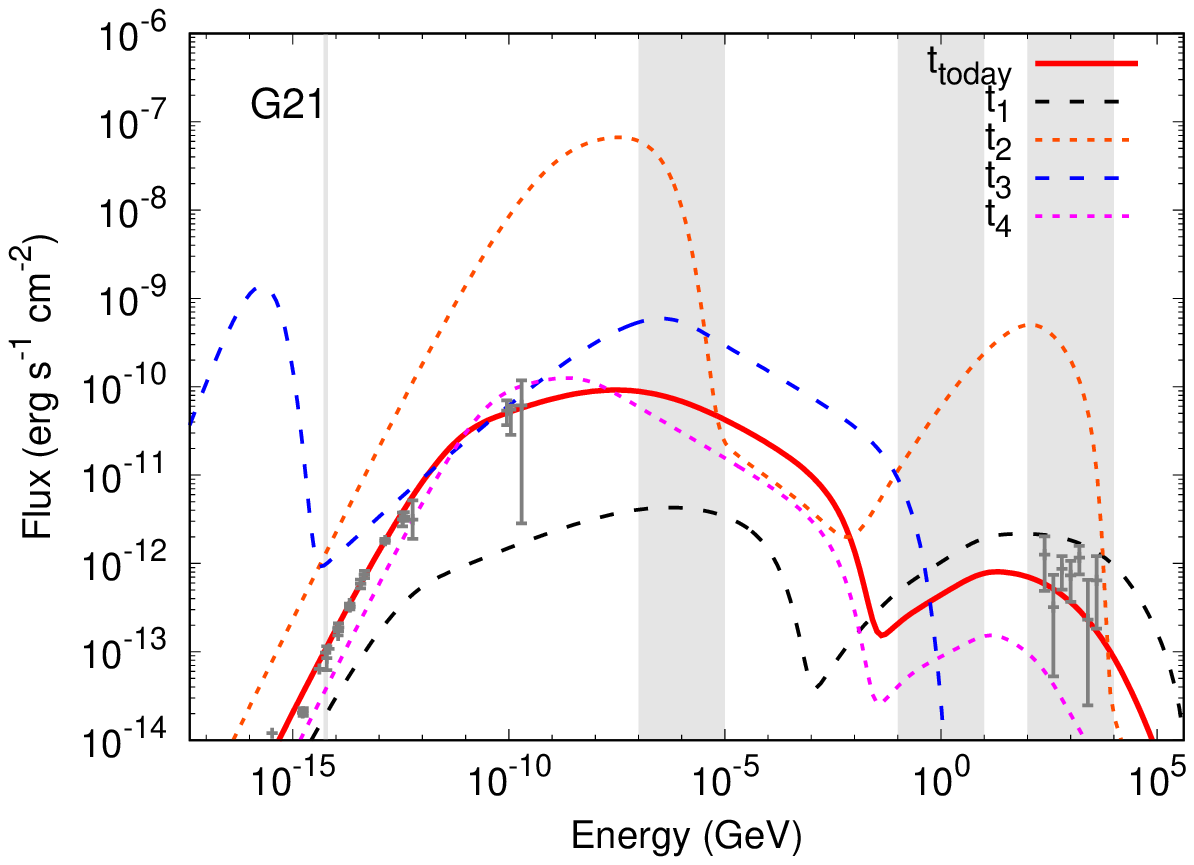} \\
\hspace{-0.5cm}
        \includegraphics[width=0.35 \textwidth]{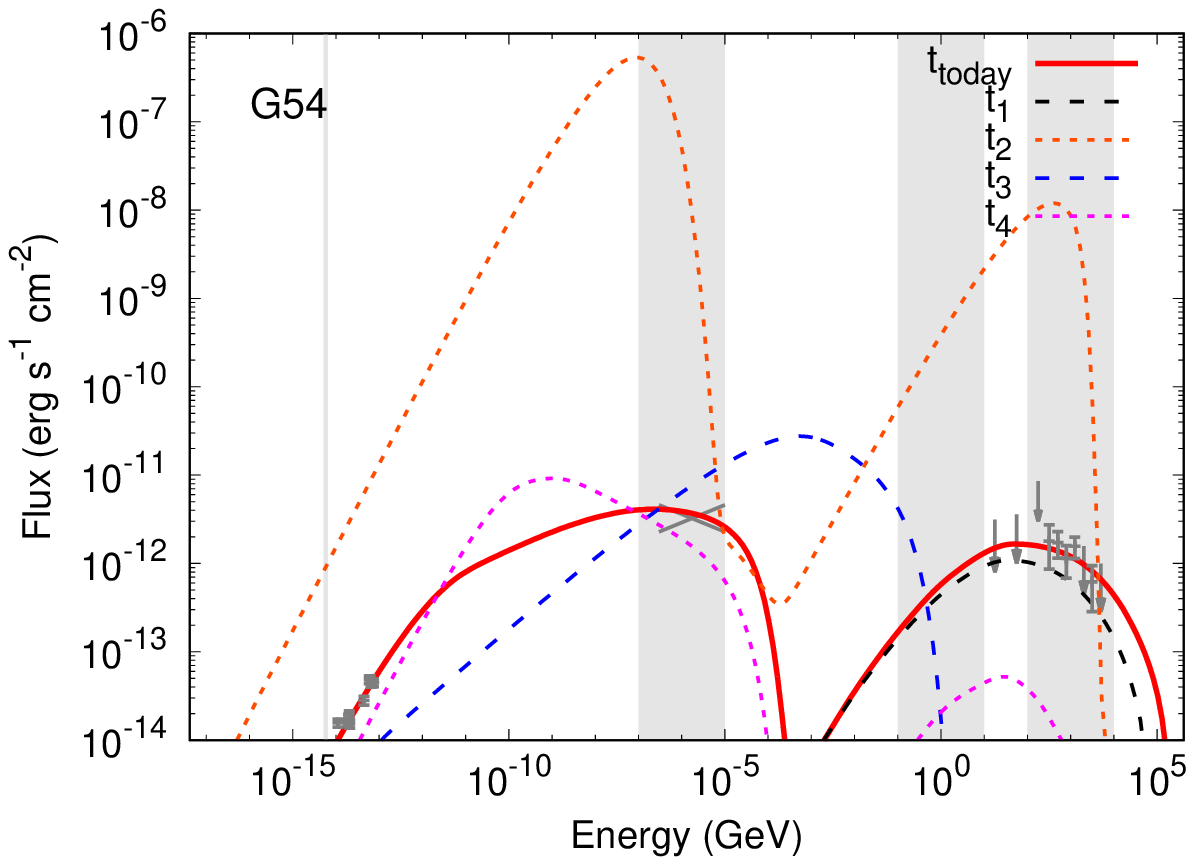}\hspace{-0.4cm}
        \includegraphics[width=0.35 \textwidth]{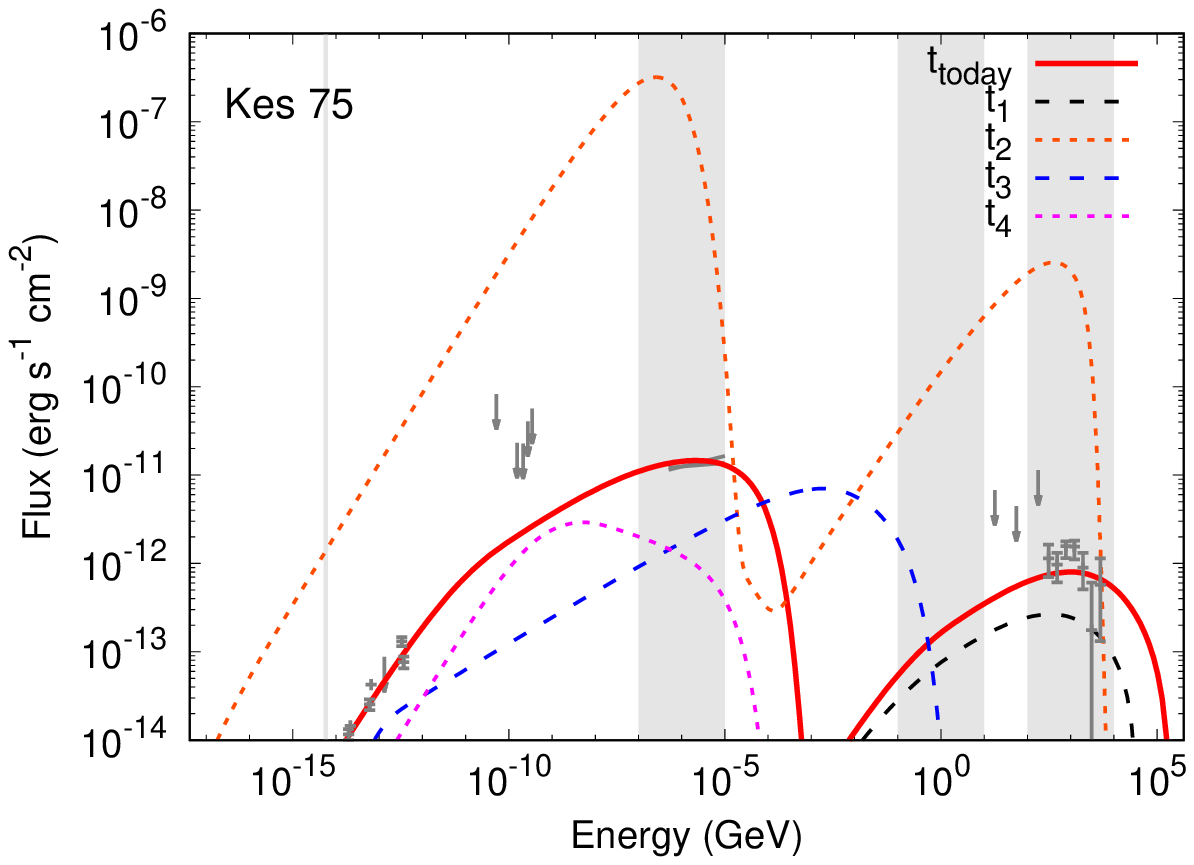} \hspace{-0.5cm}
                \includegraphics[width=0.35 \textwidth]{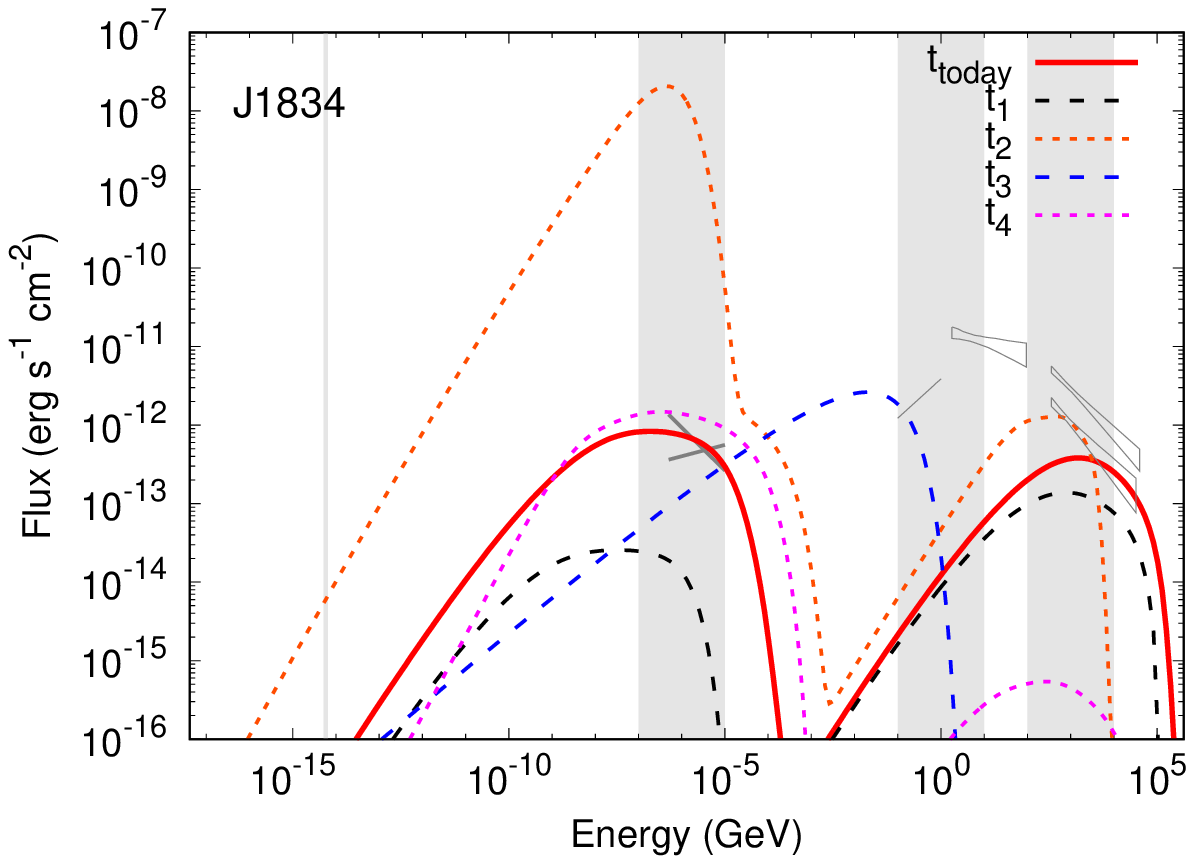}\\
\hspace{-0.5cm}                
    \includegraphics[width=0.35 \textwidth]{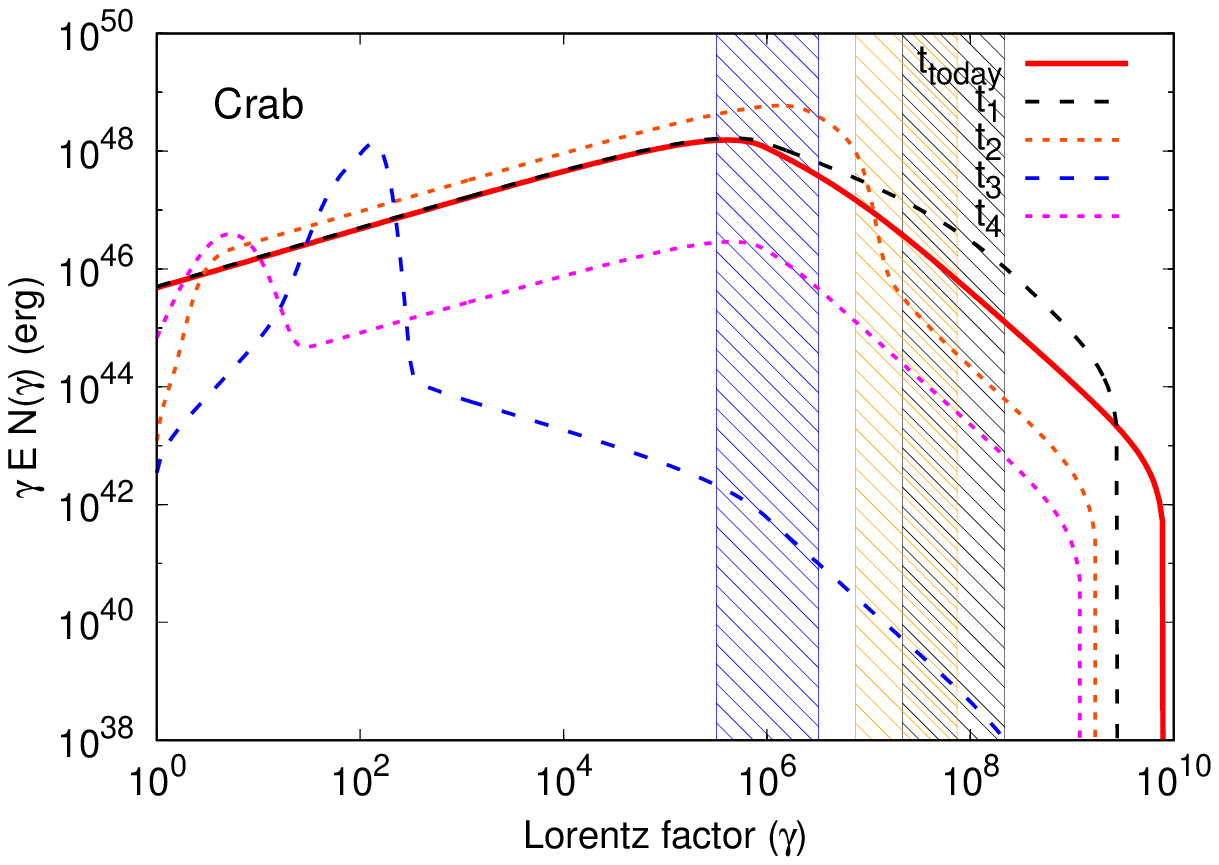}\hspace{-0.4cm}
    \includegraphics[width=0.35 \textwidth]{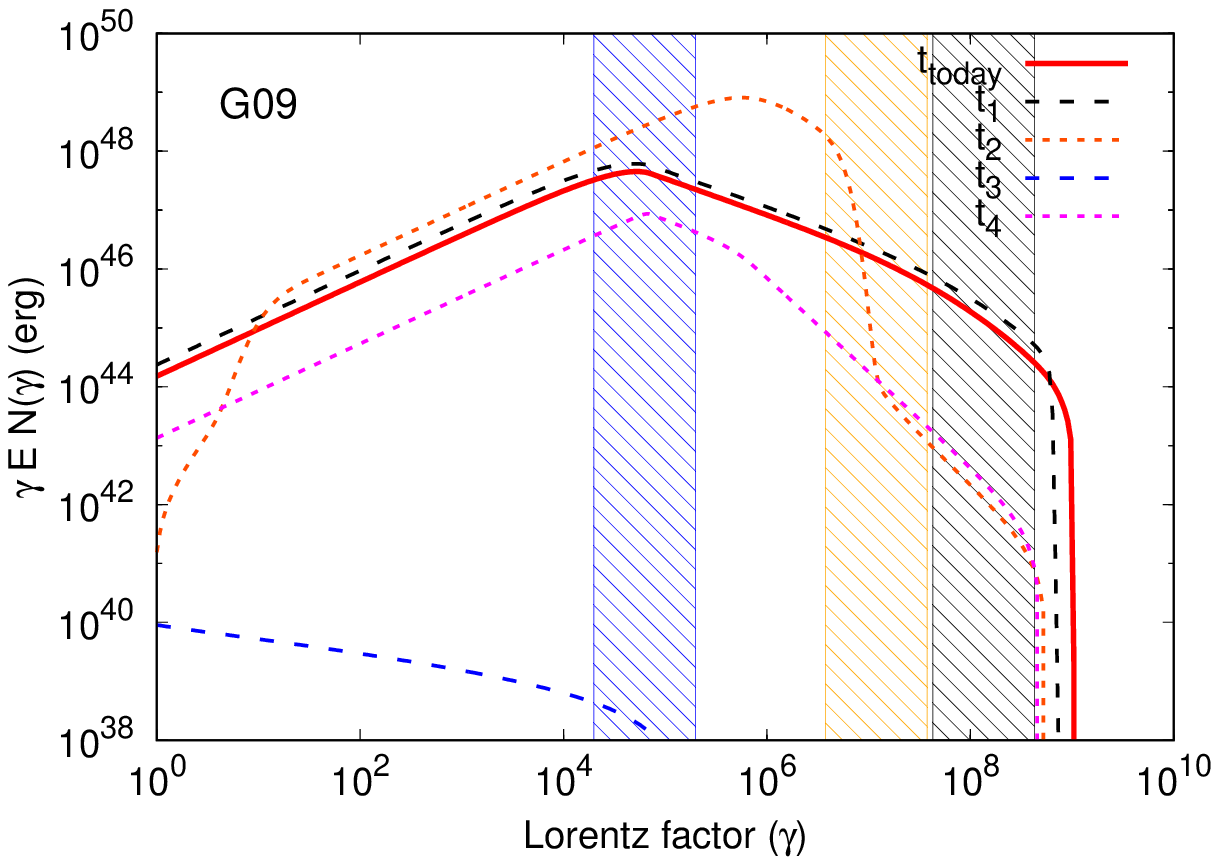}\hspace{-0.4cm}
    \includegraphics[width=0.35 \textwidth]{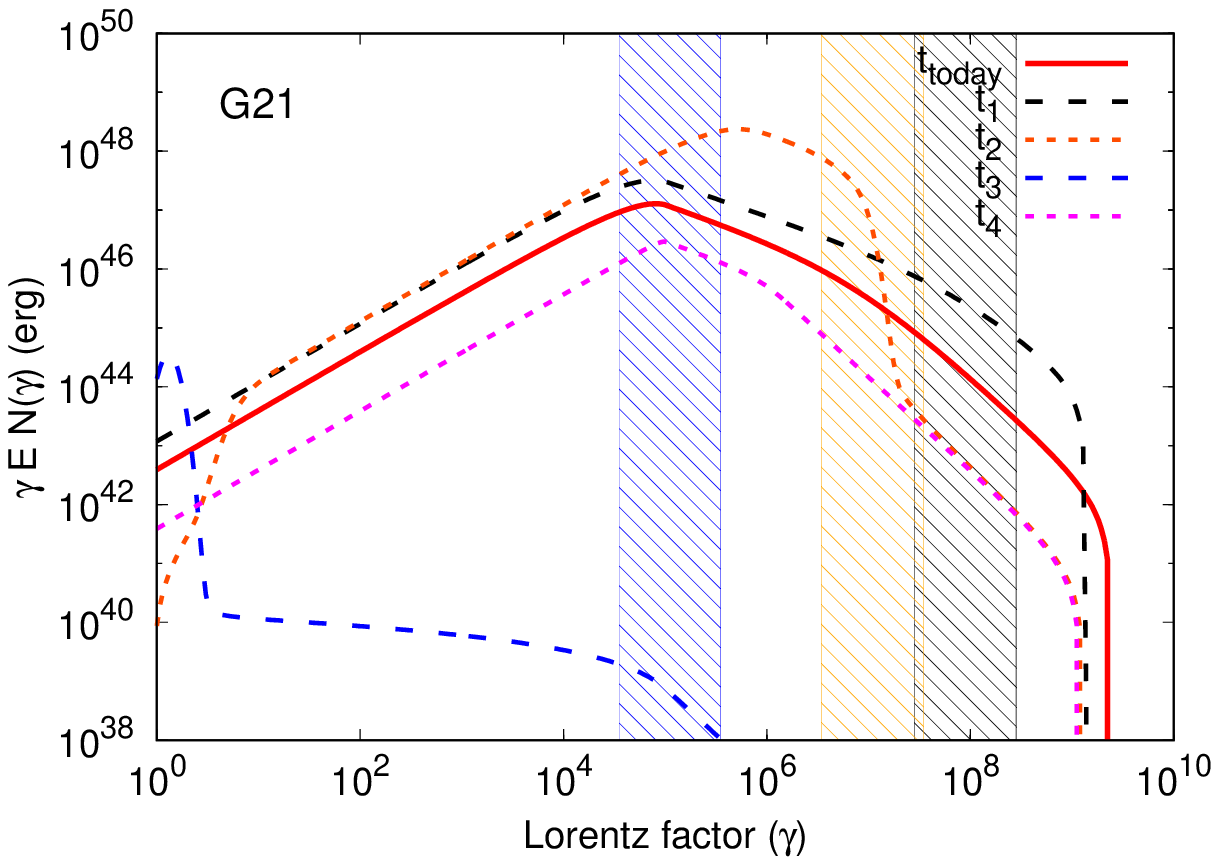}\\
\hspace{-0.5cm}
         \includegraphics[width=0.35 \textwidth]{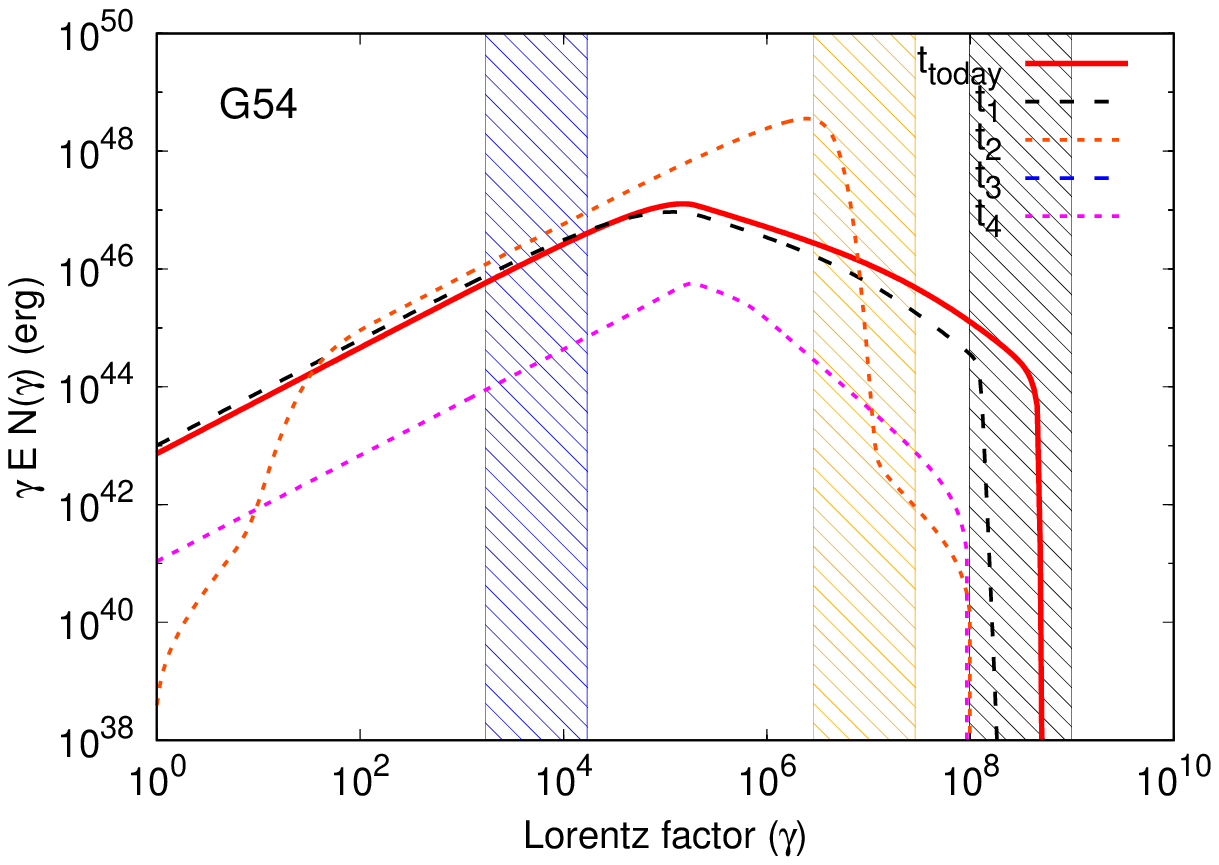}\hspace{-0.4cm}
    \includegraphics[width=0.35 \textwidth]{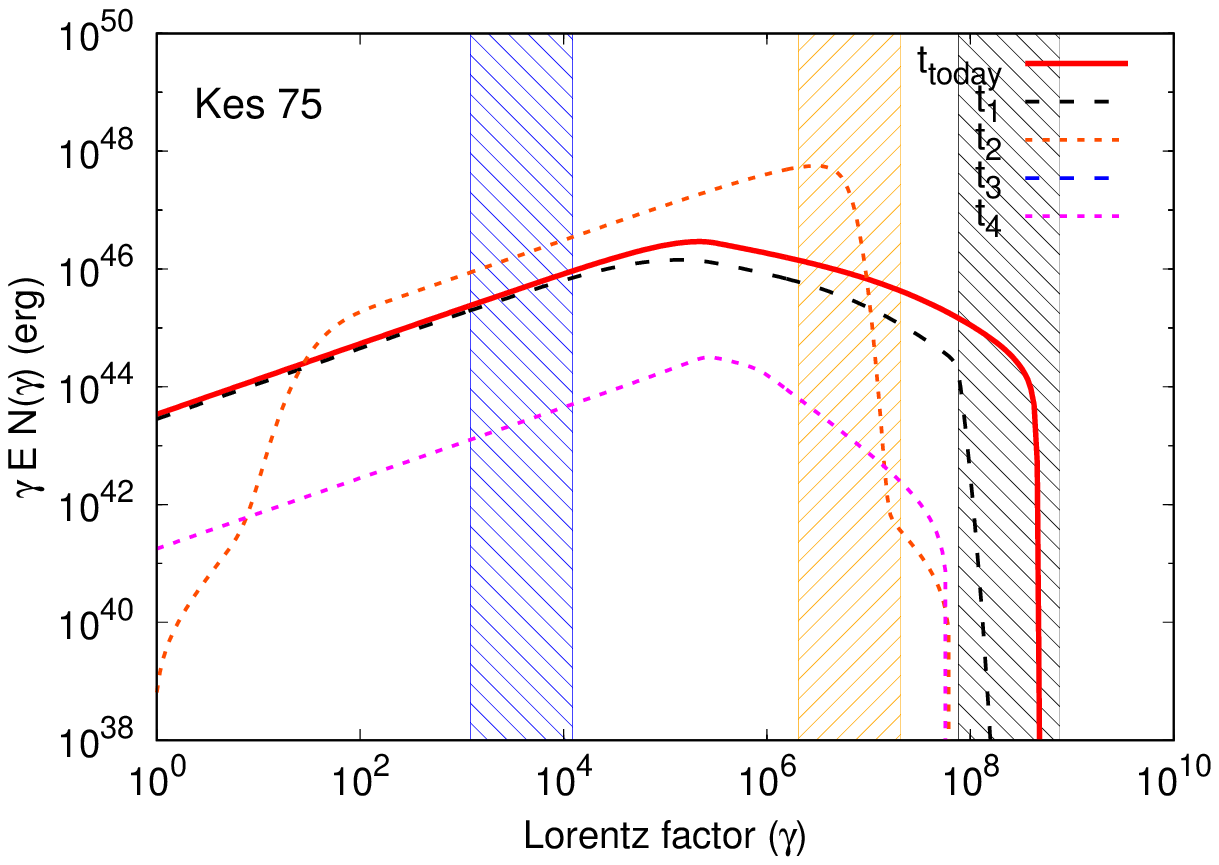}\hspace{-0.4cm}
    \includegraphics[width=0.35 \textwidth]{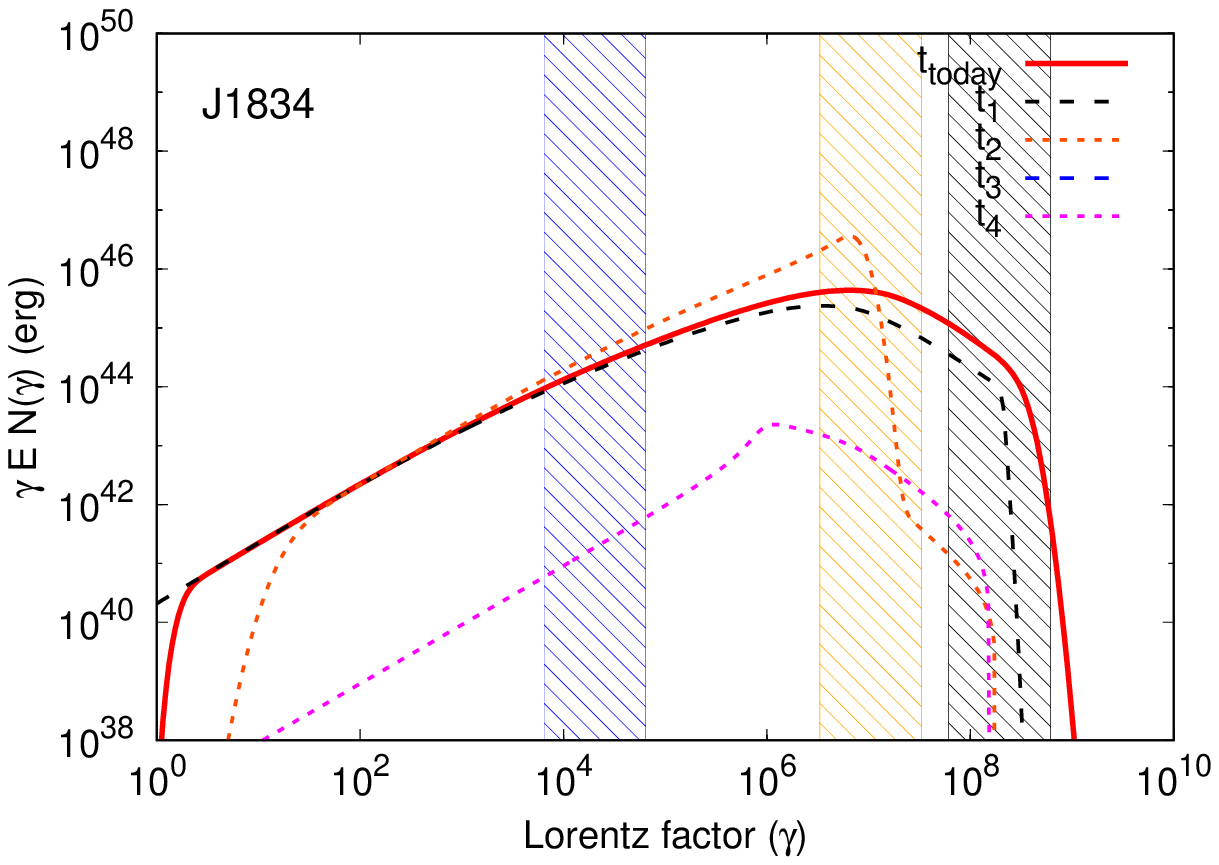}   \\
    \caption{Spectral energy and electron distributions of the modelled PWNe along time. Each panel shows the evolution at different moments of interest for each nebula ($t_1 \ldots t_4)$, which are introduced and discussed in the text, and includes also the results at the age today for comparison. The colored shadows in the electron panels note the Lorentz factor whose synchrotron-emitted characteristic energy is in the X-ray band (0.1--10 keV) for the nebular magnetic field value at $t_1$ (black), $t_2$ (orange) and $t_3$ (blue), respectively. The shadows in the SEDs note the radio (1.4 GHz), X-rays (0.1-10 keV), GeV (0.1-10 GeV), and TeV (1-10 TeV) bands used to compute the corresponding luminosities.
}
    \label{time_evolution}
\end{figure*}

The theoretical approach described is able to cope well with multifrequency data of known nebulae. 
The red curves in the top panels of 
Fig. \ref{time_evolution} shows the spectral energy distribution (SEDs) 
of the six PWNe (Crab, G09, G21, G54, Kes75, and J1834) that we take as examples in this work, at their corresponding age today as fixed or deduced from observations. 
The parameters for each model, together with the relevant 
pulsar's observational data are given in Table 1. 
Notation for all the parameters follows that usually found in the literature, and in any case, is consistent with that used 
by us before \citep{Martin2016,Torres2017}.
We divide parameters in Table 1 among measured or assumed,
derived, and fitted values. Apart of these parameters we assume the following usual ones for all PWNe/SNR complexes: energy of the explosion
$E_{SN}=10^{51}$ erg, interstellar medium density $\rho_{ISM}=0.5 $ cm$^{-3}$,
 SNR density index  = 9,
 PWN adiabatic index    = 1.333, and
 SNR adiabatic index    = 1.667.
We also consider the cosmic microwave background  with
  $T_{cmb} =        2.73$ K and
 $\omega_{cmb}=0.25$ eV cm$^{-3})$.
As expected, small variation in the fitted parameters 
are found when compared with similar models but that do not take into account 
reverberation \cite{Torres2014}.

Note that all PWNe studied are now relatively young, and considered to be free-expanding except for J1834.
All other nebulae, such as Crab itself, will enter into reverberation sometime in their future.
We choose these young nebulae (rather than other more mature) 
on purpose:  as we shall see, reverberation is a very sensitive process, 
leading to a strong evolution where most of the electron population is wiped out. 
Since we are actually interested in the reverberation process itself, 
fixing the model parameters before this process happens makes more sense than doing it long after it ends.

%%%%%%%%%%%%%%%%%%%%%%%%%%%%%%%%%%%%%%%%%%%%%%%%%%
%\section{Time evolution}
%%%%%%%%%%%%%%%%%%%%%%%%%%%%%%%%%%%%%%%%%%%%%%%%%%

Fig. \ref{time_evolution} shows the time evolution of the model fitting the current data for each of the PWNe considered. 
The two sets of panels show
the electron and spectral energy distributions along time. 
A strong time evolution is expected. 
The times shown are chosen within and around the corresponding reverberation period of each PWN,  
and correspond to 
the times of the maximum PWN radius,
$t_1=t(R_{max})$,
the maximum of the X-ray efficiency,
$t_2=t($Eff$^{max}_{X})$,
the minimum PWN radius,
$t_3=t(R_{min})$,
and a later time already at the Sedov phase,
$t_4=t(@Sedov)$.
Specific values for these times along the PWNe evolution of each nebula are also given in Table 1.
The X-ray efficiency (and correspondingly, radio, GeV, and TeV efficiencies as well) are defined as the ratio
of the luminosity emitted in a given frequency range at a given time with respect to the spin-down power at that same time,
e.g., Eff.$_X (t) = L_X(t) / L_{sd}(t)$. If at a time $t$ we measure this ratio to be larger than 1, we shall say the PWN is superefficient.

%%%%%%%%%%%%%%%%%%%%%%%%%%%%%%%%%%%%%%%%%%%%%%%%%%
\section{Superefficiency}
%%%%%%%%%%%%%%%%%%%%%%%%%%%%%%%%%%%%%%%%%%%%%%%%%%

\begin{figure*}
    \centering
\hspace{-0.5cm}
    \includegraphics[width=0.35 \textwidth]{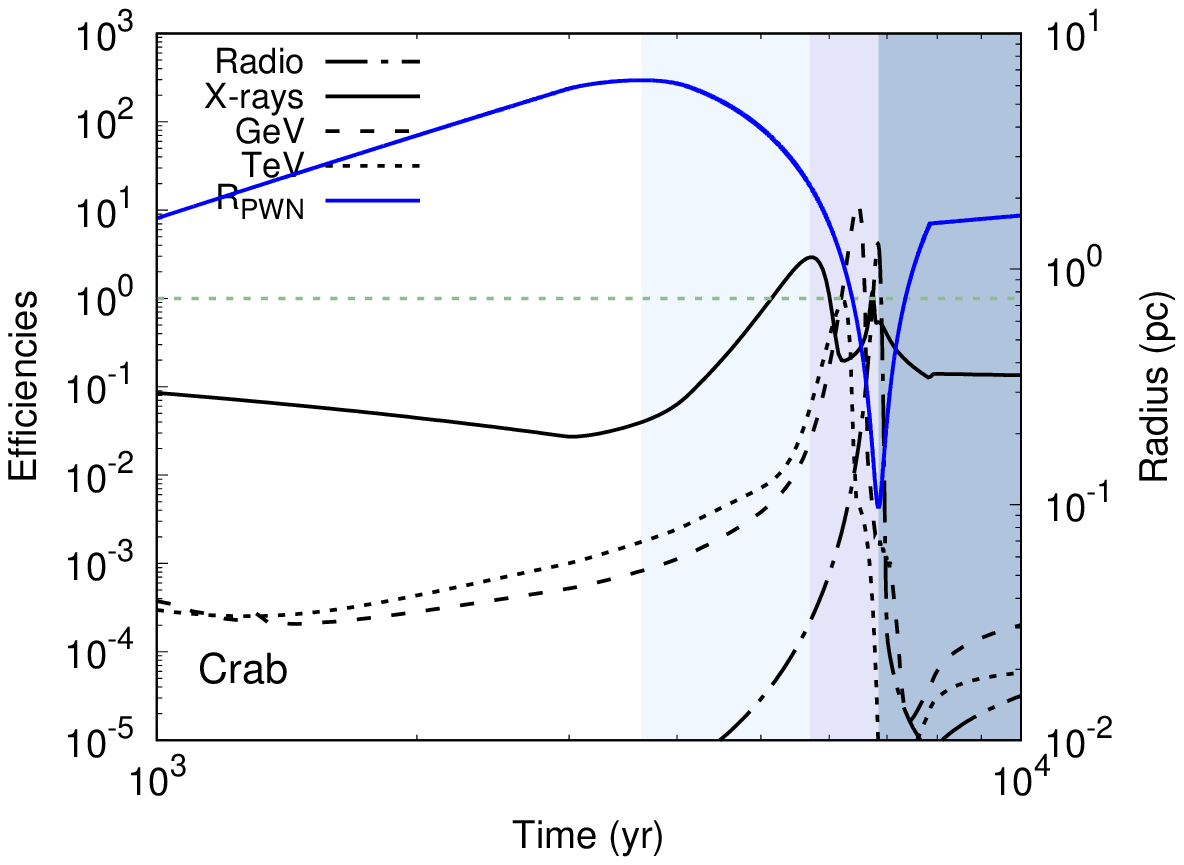}\hspace{-0.4cm}
    \includegraphics[width=0.35 \textwidth]{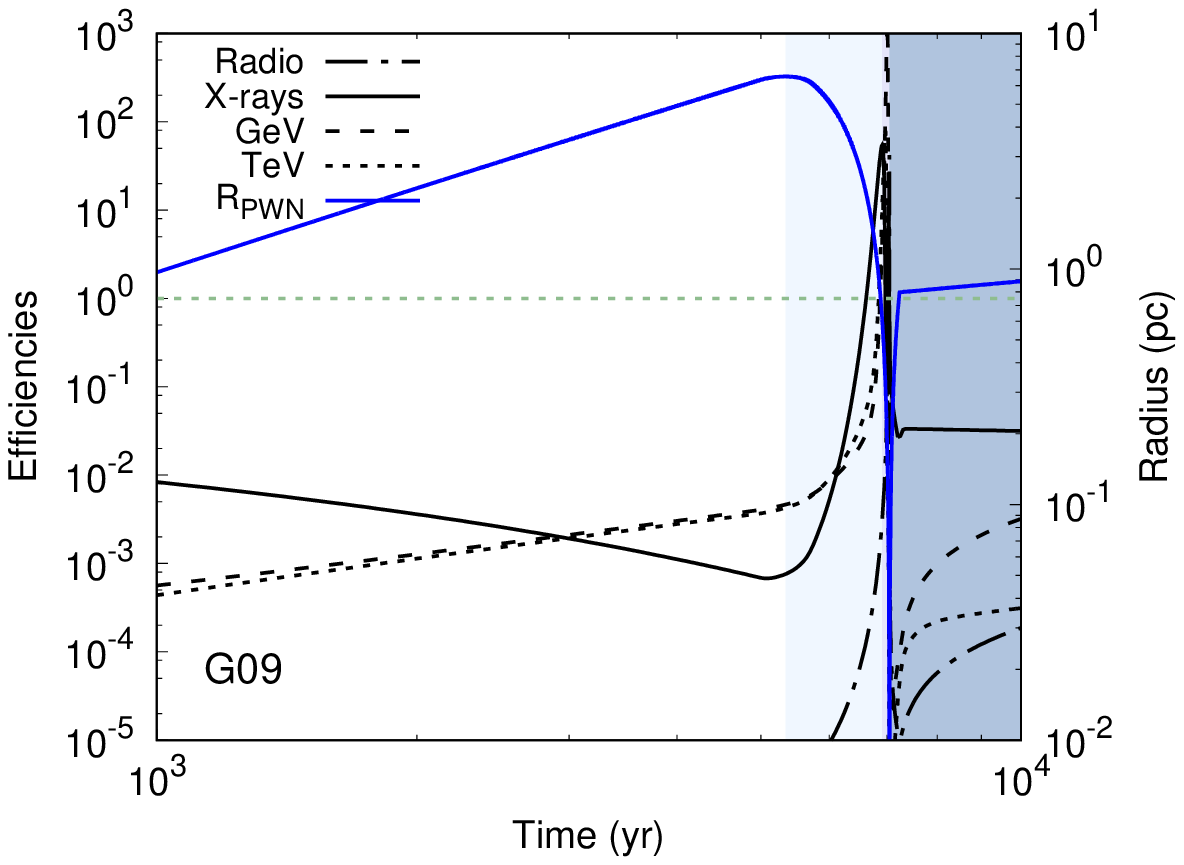}\hspace{-0.4cm}
    \includegraphics[width=0.35 \textwidth]{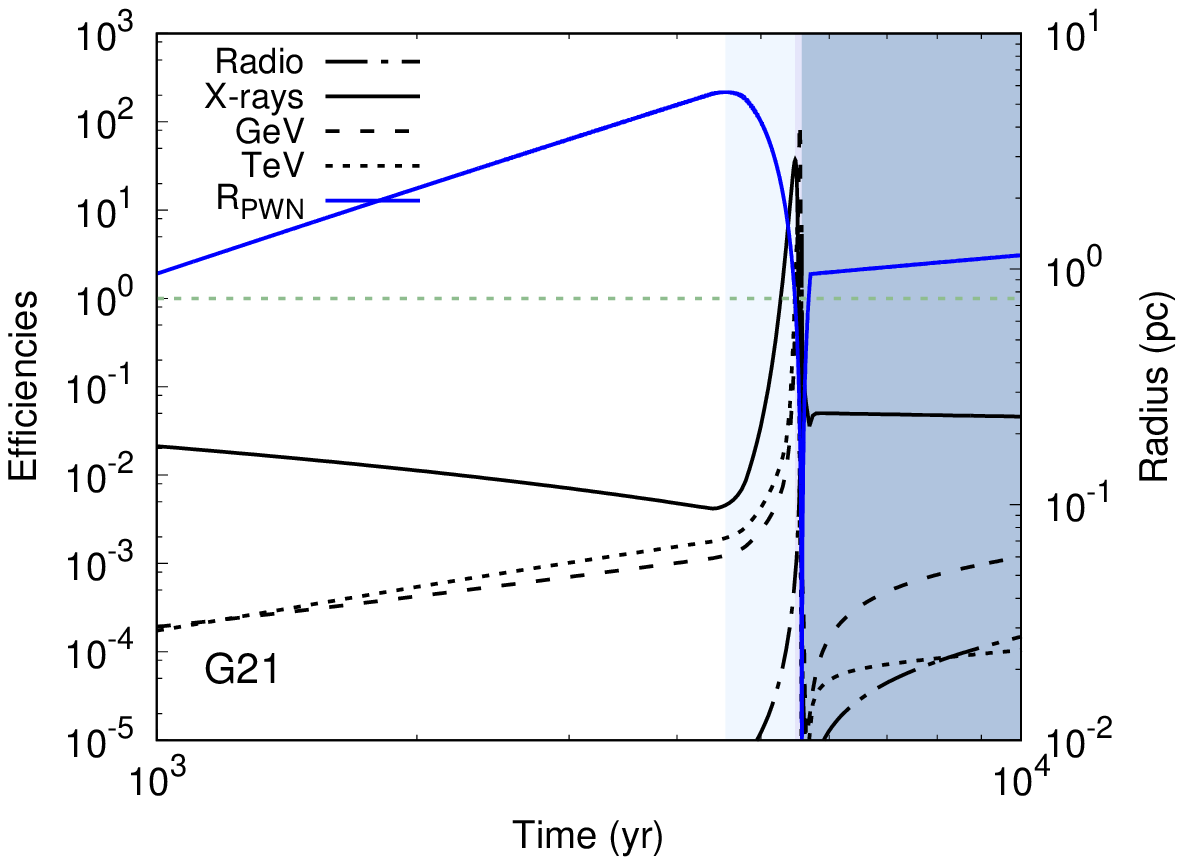}\\
\hspace{-0.5cm}
         \includegraphics[width=0.35 \textwidth]{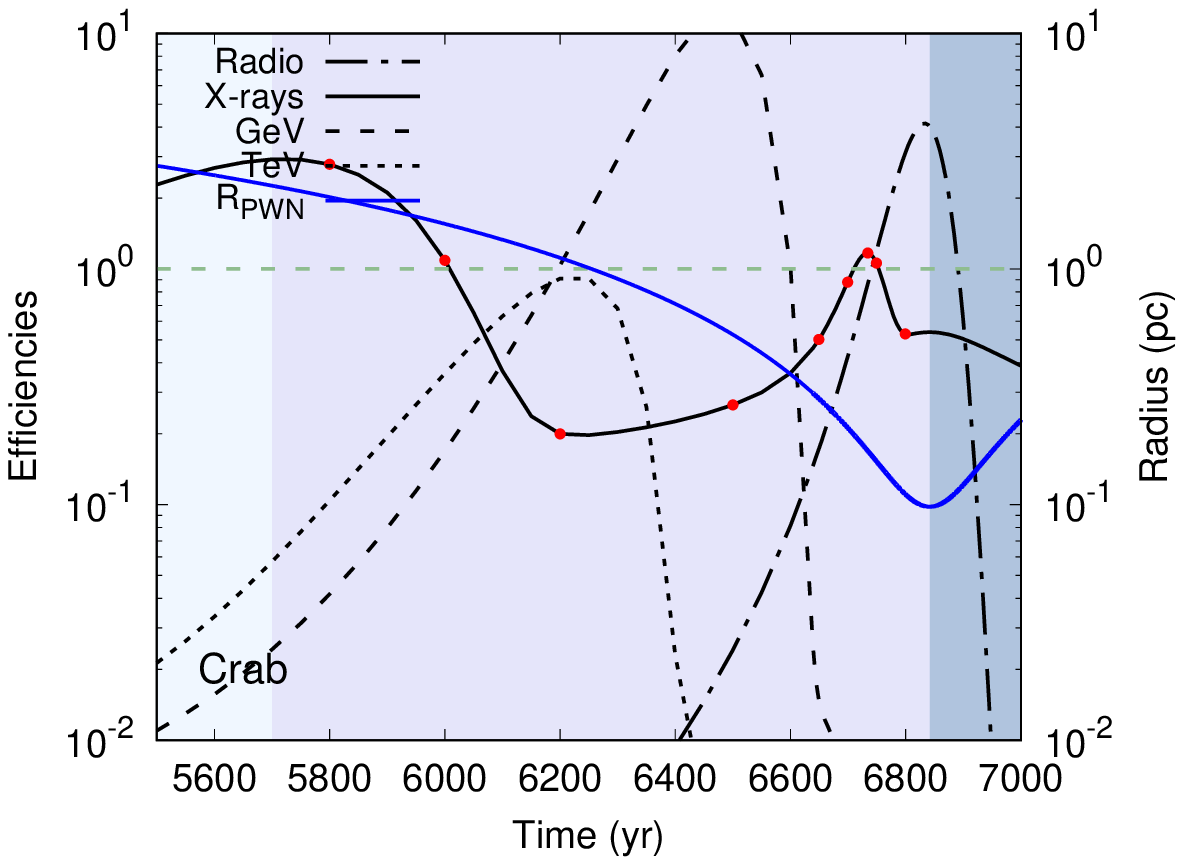}\hspace{-0.4cm}
    \includegraphics[width=0.35 \textwidth]{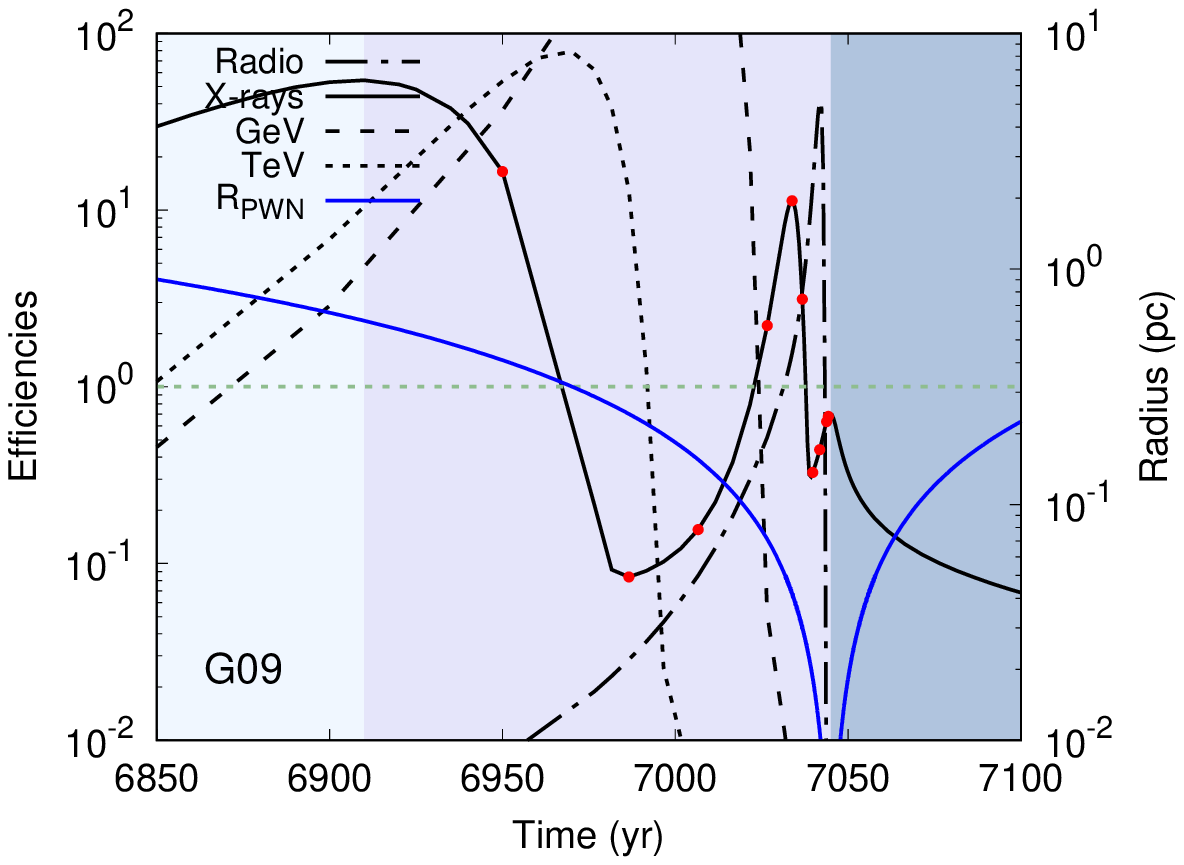}\hspace{-0.4cm}
    \includegraphics[width=0.35 \textwidth]{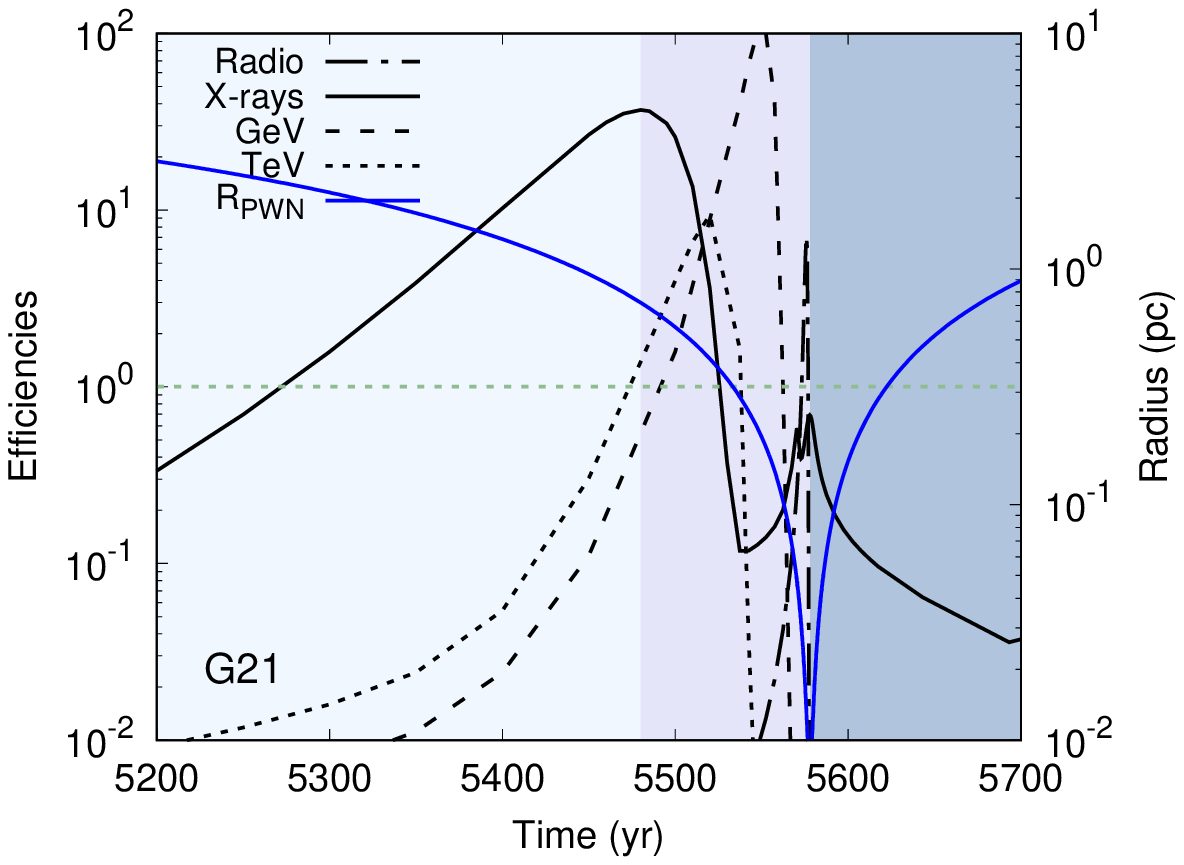}   \\
 \hspace{-0.5cm}
            \includegraphics[width=0.35 \textwidth]{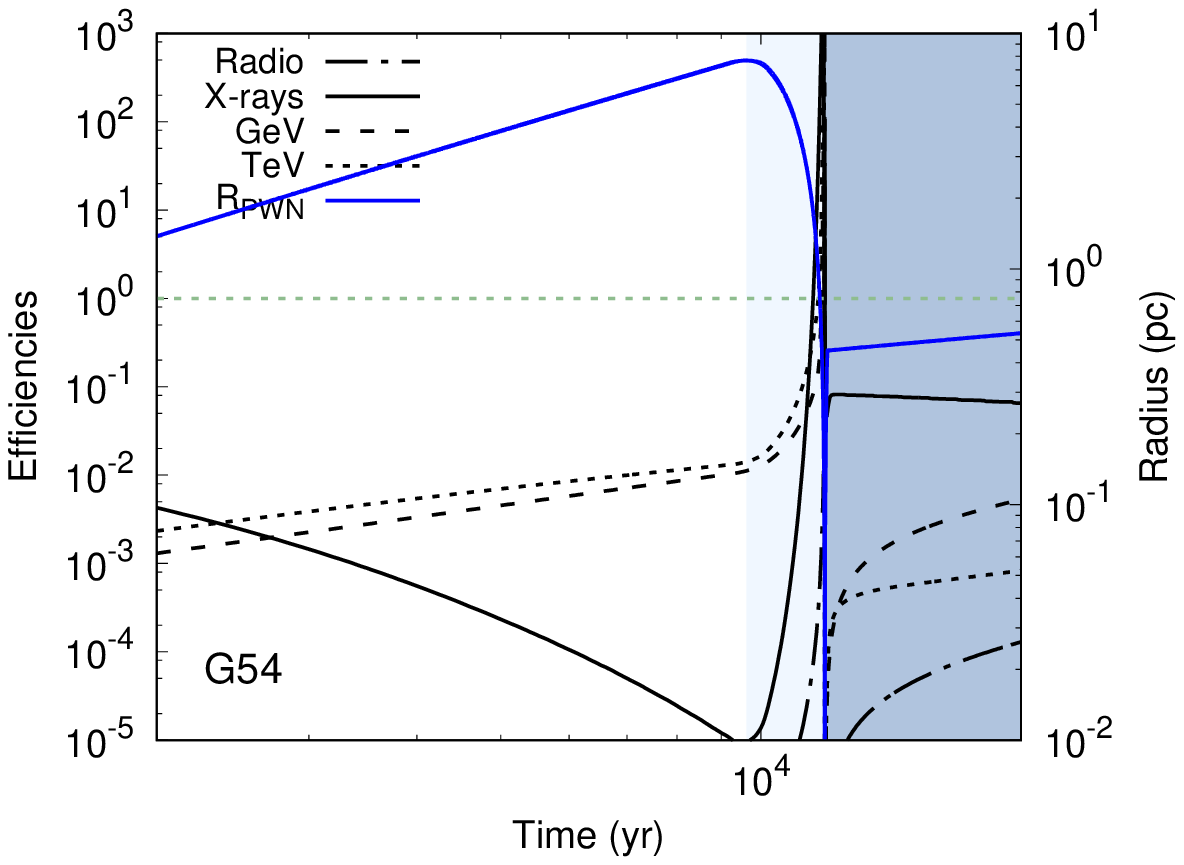}\hspace{-0.4cm}
        \includegraphics[width=0.35 \textwidth]{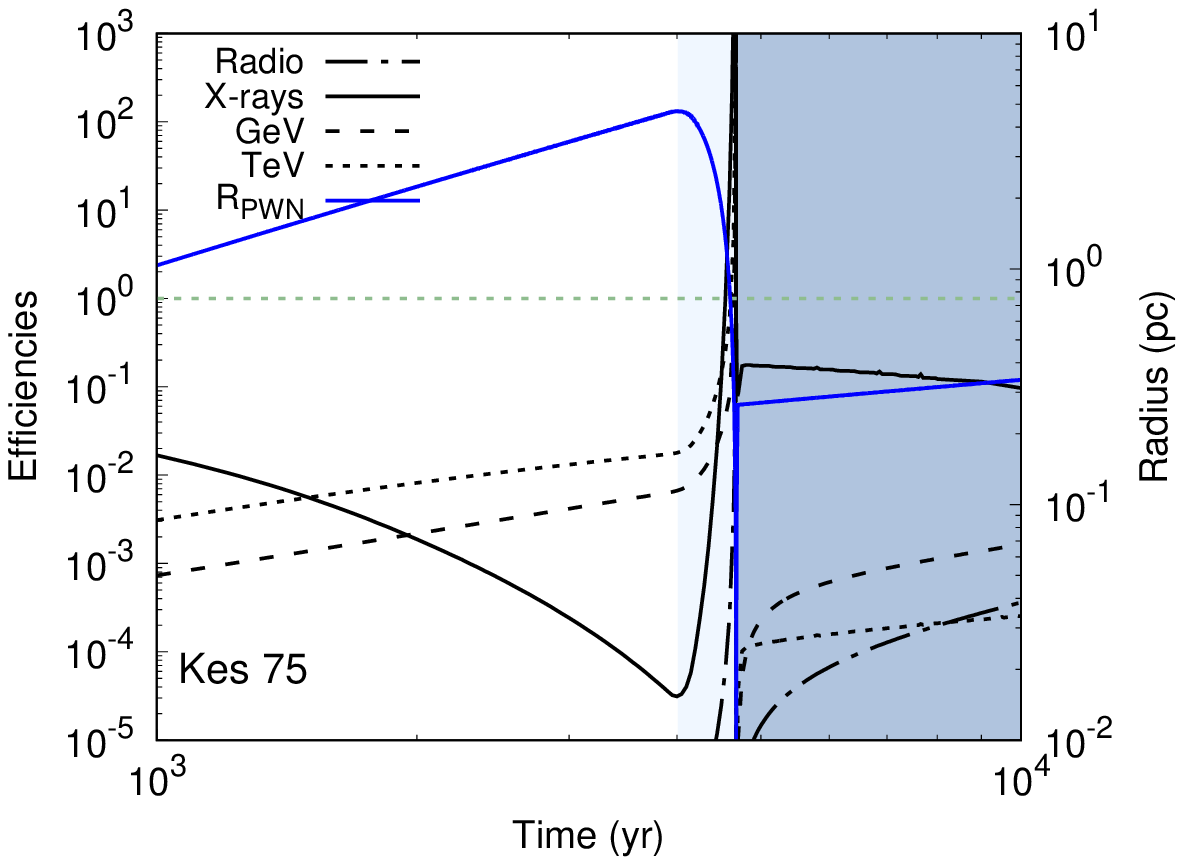} \hspace{-0.5cm}
       \includegraphics[width=0.35 \textwidth]{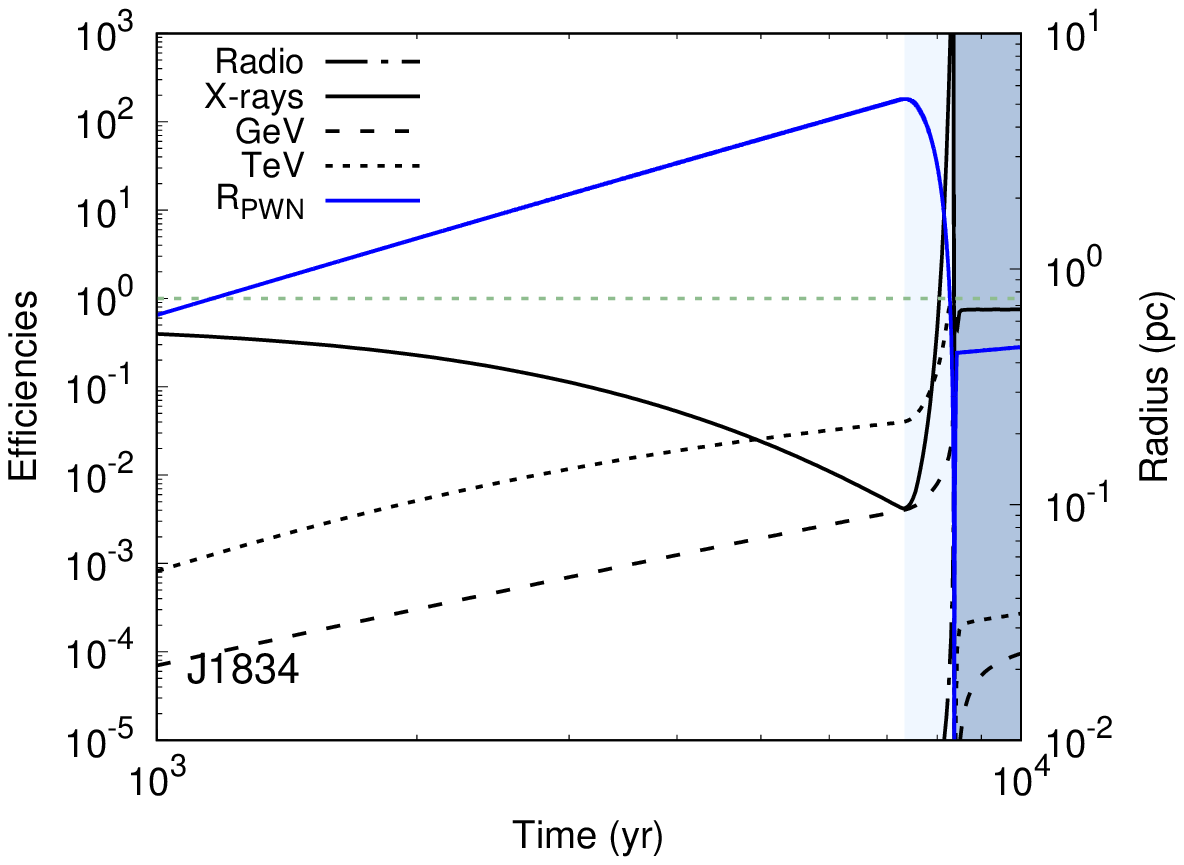}\\
\hspace{-0.5cm}
        \includegraphics[width=0.35 \textwidth]{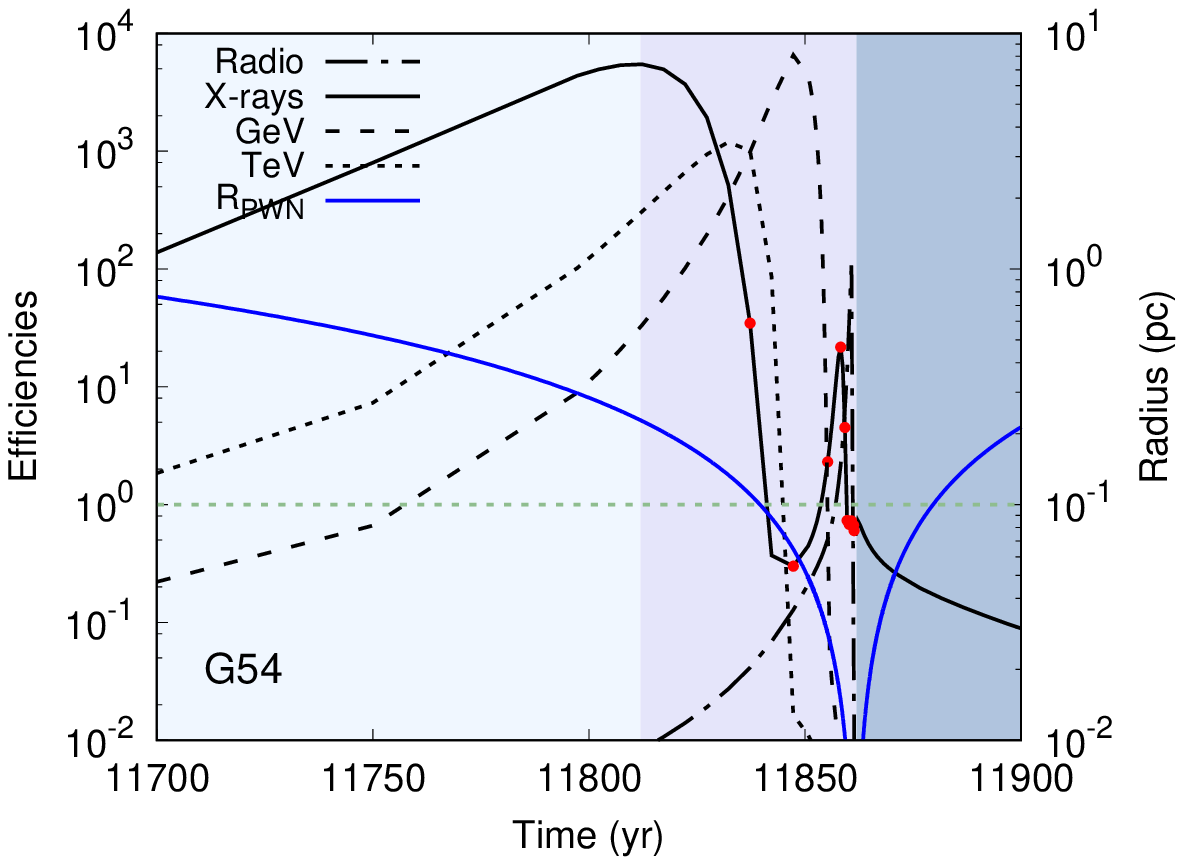}\hspace{-0.4cm}
        \includegraphics[width=0.35 \textwidth]{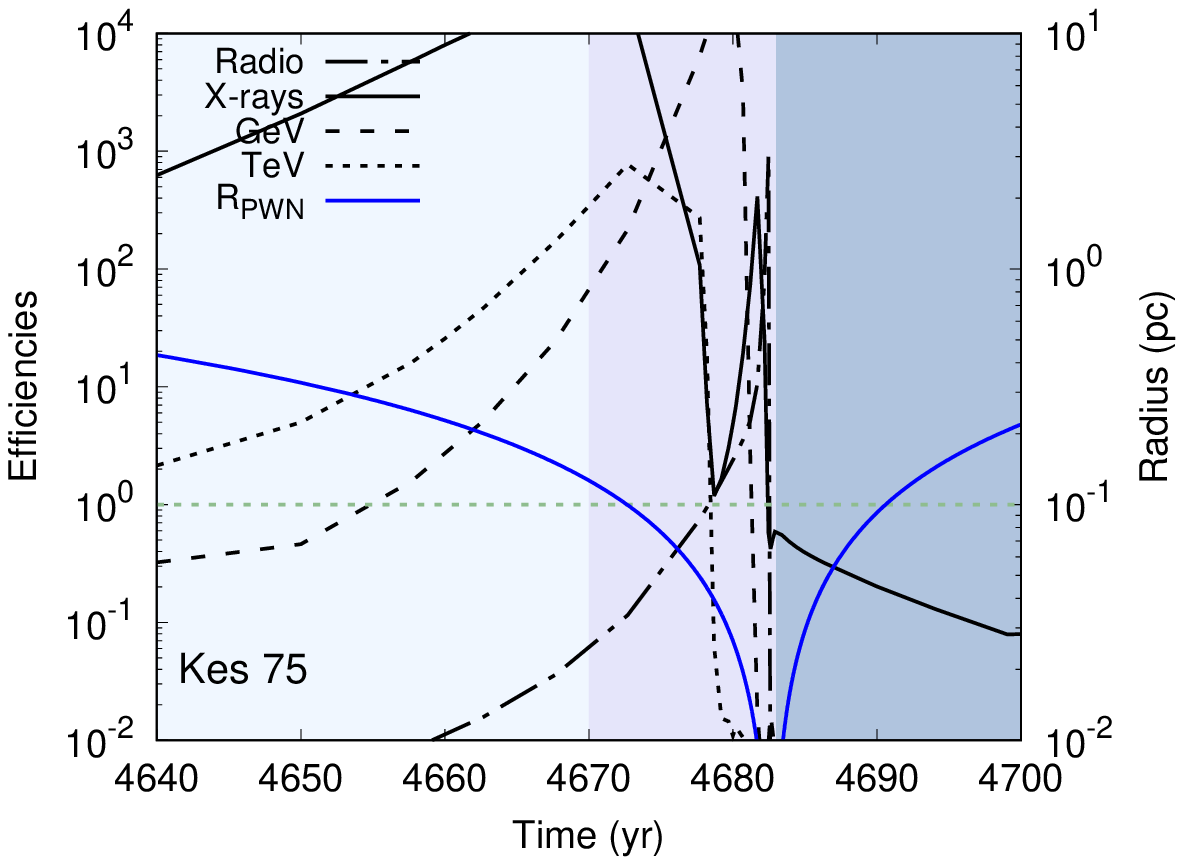} \hspace{-0.5cm}
\includegraphics[width=0.35 \textwidth]{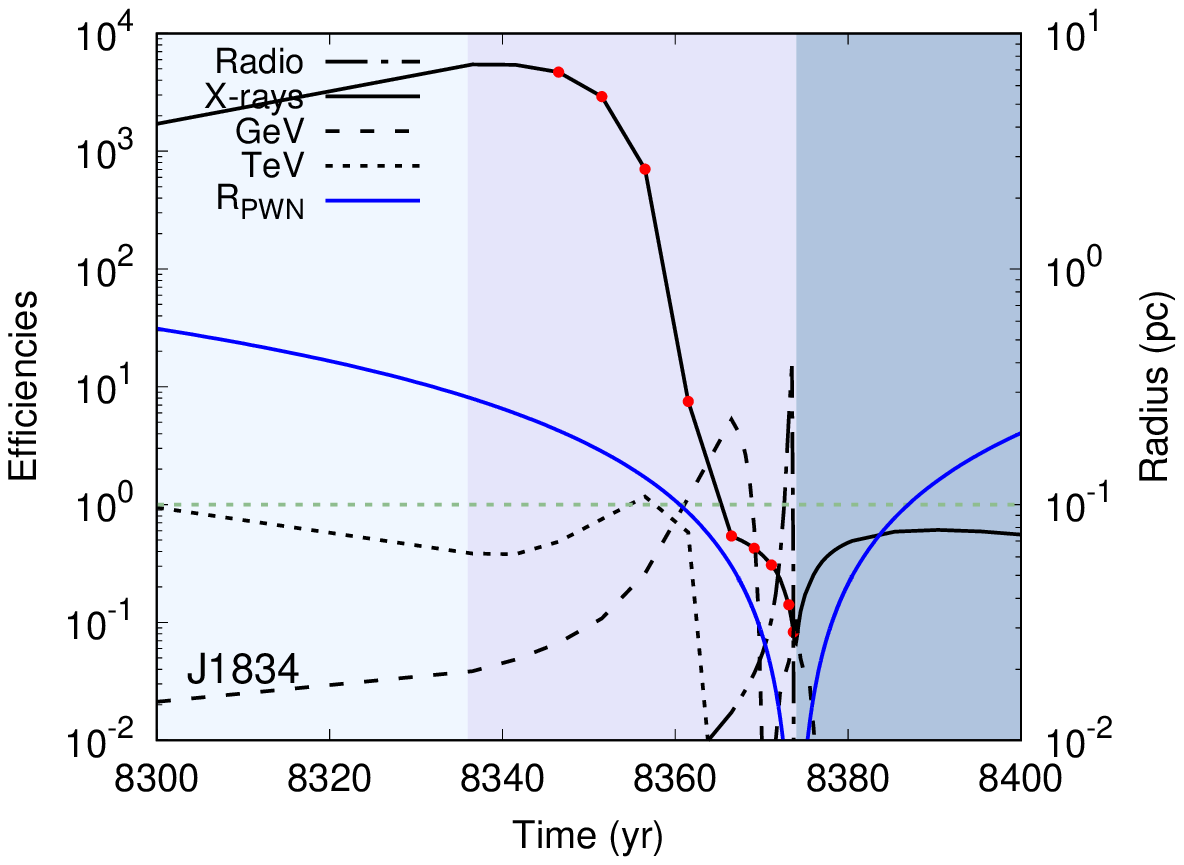}
    \caption{Evolution of the PWNe efficiencies in X-rays (0.1-10 keV), GeV (0.1-10 GeV), and TeV (1-10 TeV) and PWN radii along time. The second and fourth rows zooms around the reverberation period, as shown in the corresponding global evolution panels. 
  }
    \label{super}
\end{figure*}

Fig. \ref{super} shows the time evolution of the calculated efficiencies
 in radio (1.4 GHz), X-rays (0.1-10 keV), GeV (0.1-10 GeV), and TeV (1-10 TeV), together with the PWN radii. 
Table 1  shows the 
 timescale for the duration of reverberation ($t(R_{min})-t(R_{max})$), the minimum radius, and the maximum magnetic field attained, as well as the properties of any
superefficiency period in radio, X-rays, GeV,  or TeV energies (maximum efficiency, Eff.$^{max}$; duration, Dos; and the time at which the maximum efficiencies happen, $t($Eff.$^{max}$)).
We also show the time $t_4$ in the Sedov expansion used in the figures as an example of the spectra in this regime, and values of the magnetic field at different times of interest. 
Reverberation brings a significant evolution in a short period of time. 
Plotting efficiencies rather than distributions makes this evolution more clear.

The X-ray efficiency has several stages of increase and decrease, which can be used to define different phenomenological phases. We call them phases a to c, for reference. We distinguish these phases via the following subsequent events:
Phase a has the PWN in free expansion, and lasts from the pulsar birth to the maximum of the nebula radius (at $t_1$). Phase b finishes at the maximum of the X-ray efficiency (at $t_2$). Phase c finishes at the minimum of the radius (at $t_3$).  Phase d is the Sedov expansion, assumed to continue after $t_3$.
 We use different background colours in Fig. \ref{super}  to distinguish these phases. 
 Their spectral and electron properties at these times were shown in Fig.  \ref{time_evolution}.
 Note that in some cases, phase c is too short to be visible without a zoom in the reverberation period, as shown in the second and fourth rows of 
 Fig. \ref{super}.

Fig. \ref{super}  shows that even the Crab nebula, associated to the more energetic pulsar of the sample we study, has a period in its future time evolution where, e.g., the X-ray luminosity will exceed the spin-down power at the time. 
In fact, all PWNe in our sample are expected to have a period of radio, X-ray, and GeV superefficiency, and all but Crab and J1834, will also have a period of TeV superefficiency.
The finding of superefficiency at all frequencies dramatically shows how dangerous it is to rule out a pulsar of a given spin-down power as a possible origin of nebula whose radio, X-ray, GeV, or even TeV luminosity exceeds it. 

The zoomed panels in Fig. \ref{super} show that the moments at which the maximum efficiencies are attained are close but not exactly the same at different frequencies. 
This is a natural result of having electrons of different energies contributing to the photon spectrum at different frequencies. 
The number of electrons at a given energy is in turn a result of a balance between gains (via adiabatic heating) and losses (via radiation and escape) and the peak number is attained at different times for different energies. 
We also note that there is a variety of possibilities regarding which of the maximum efficiencies is the largest. Sometimes, like the case of J1834
and Kes 75, the largest maximum efficiency occurs for the X-rays. In others, for more energetic pulsars like Crab, G09, or G21, it occurs at the GeV band.
The evolution of the radio efficiency is quite similar for all PWNe. It shows a sharp peak happening close to the time of maximum of the compression. 
In this small time scale around $t_3$,
all the PWNe studied become superefficient in radio.
However,  the maximum efficiency attained in the radio band is typically smaller than that reached at higher frequencies, see Table 1.

\begin{figure*}
    \centering
    \includegraphics[width=0.32\textwidth]{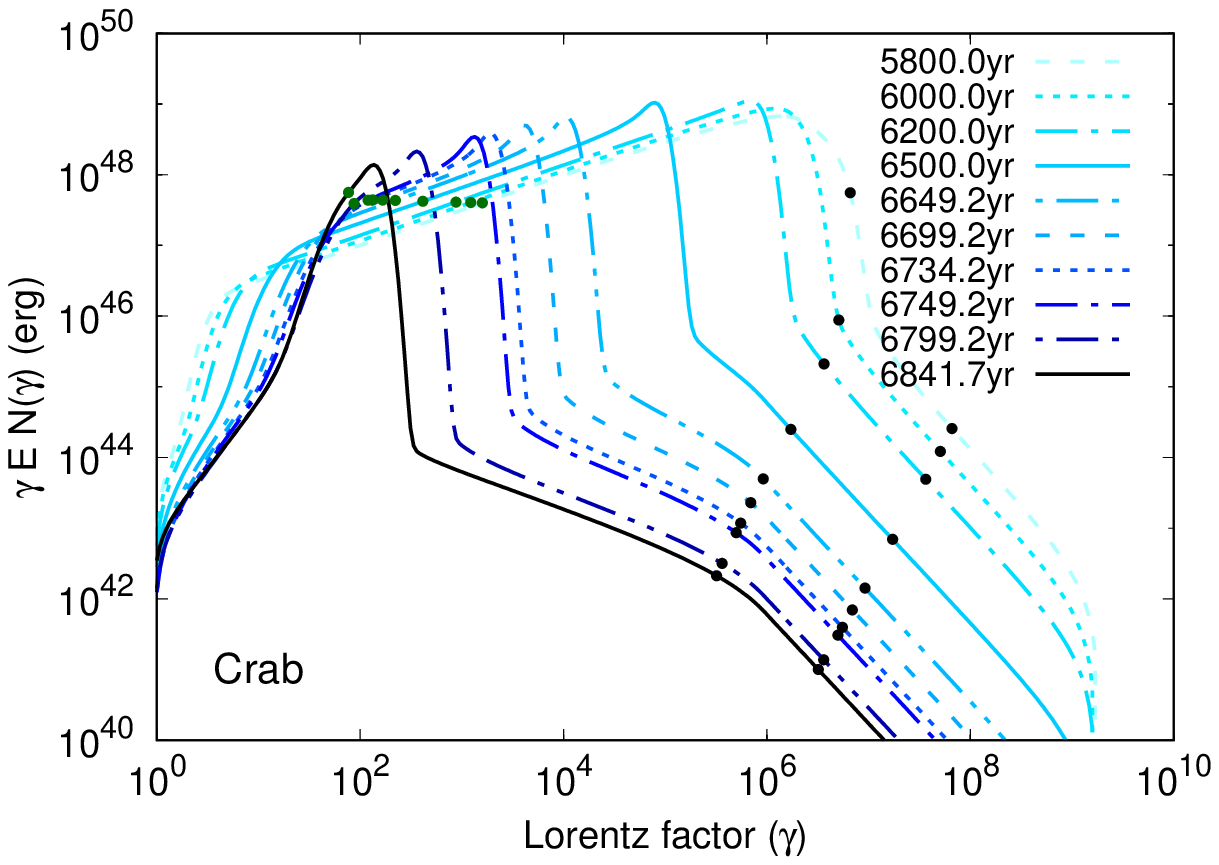}
    \includegraphics[width=0.32\textwidth]{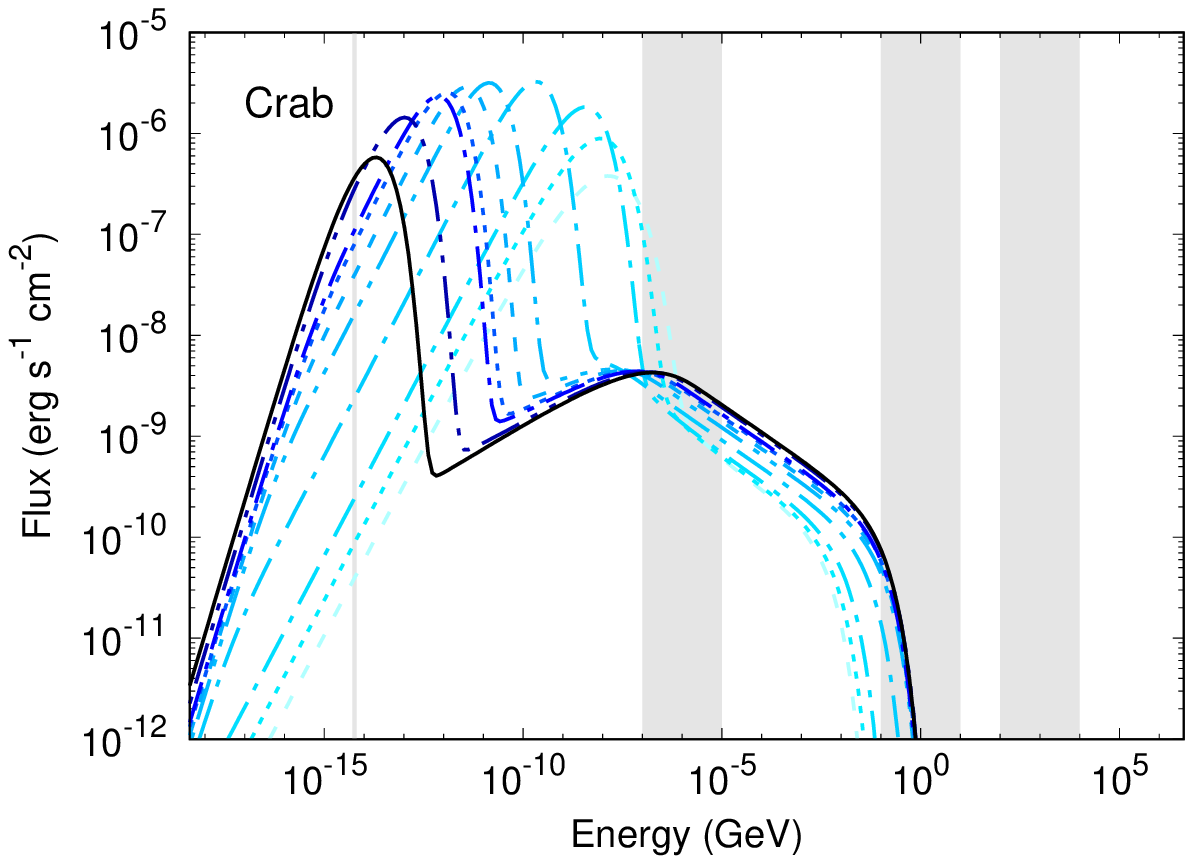}
    \includegraphics[width=0.32\textwidth]{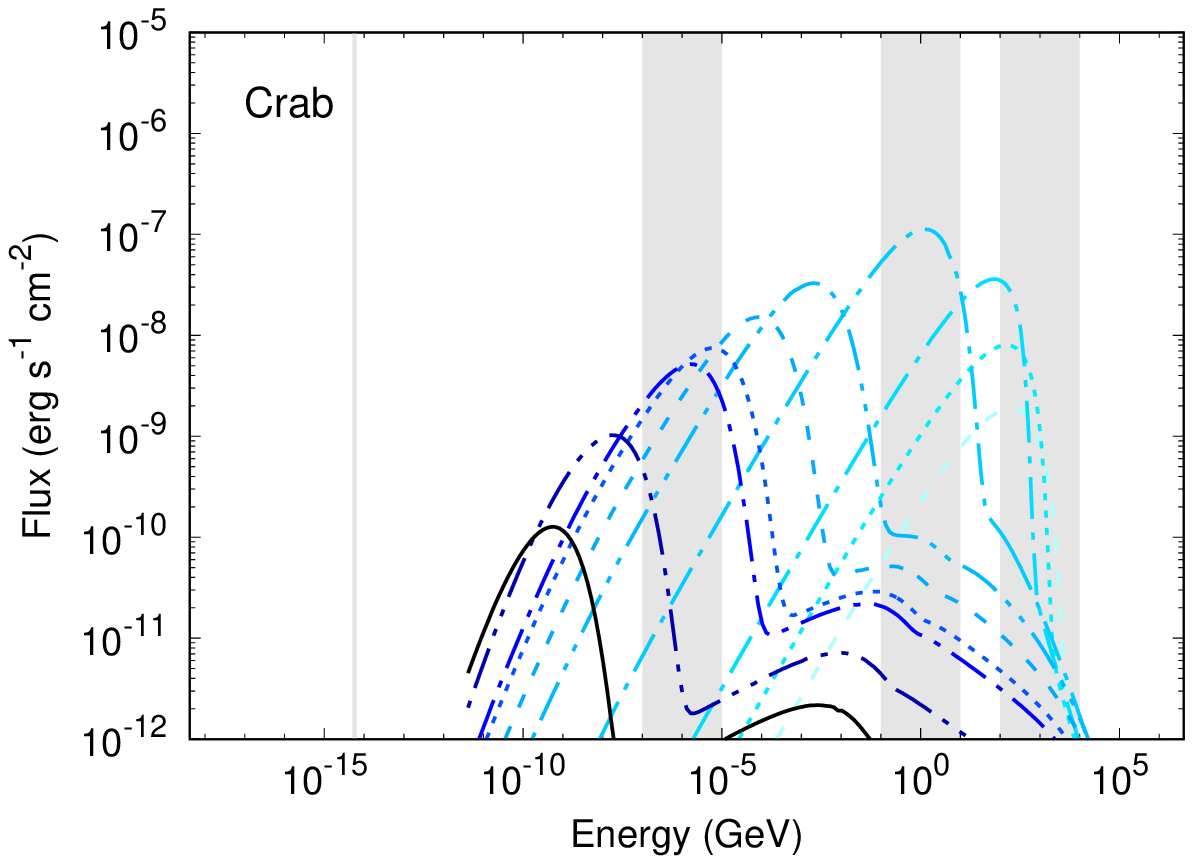}
    \includegraphics[width=0.32\textwidth]{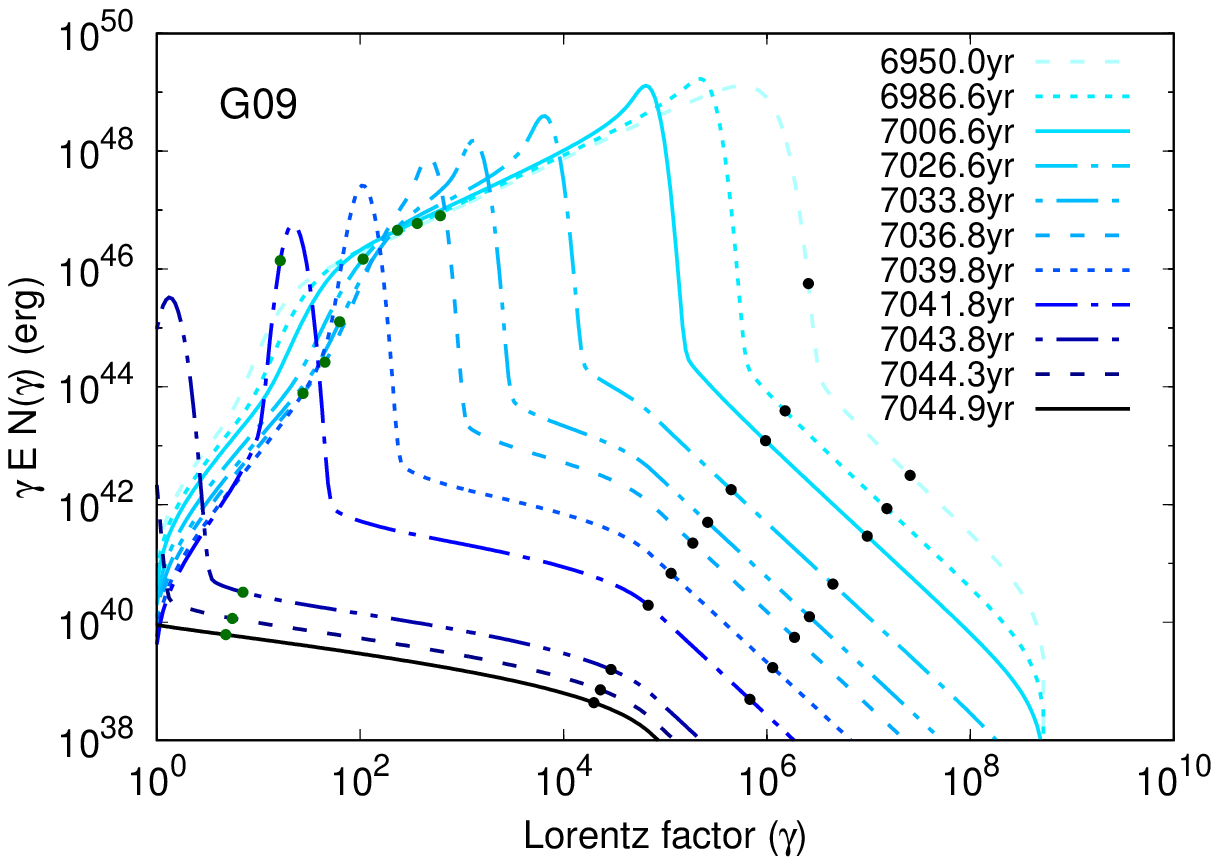}
    \includegraphics[width=0.32\textwidth]{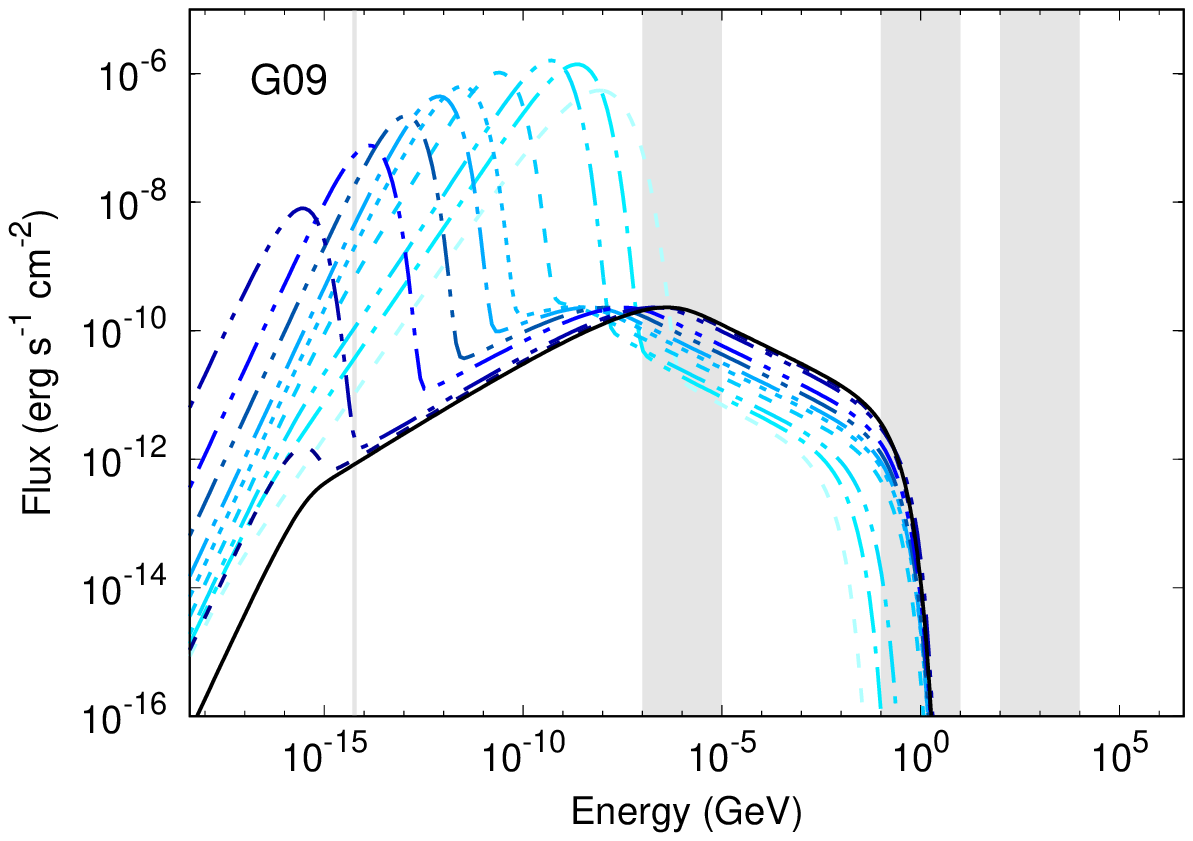}
    \includegraphics[width=0.32\textwidth]{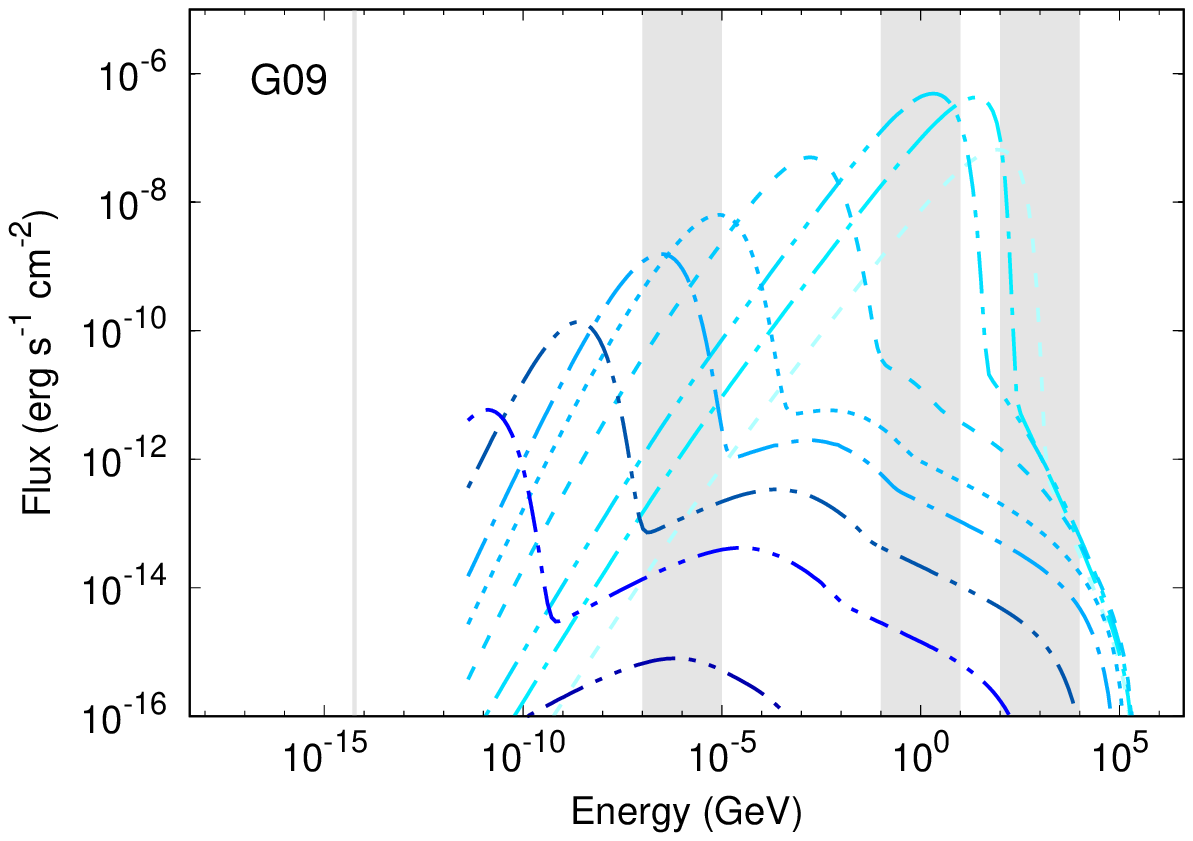}
    \includegraphics[width=0.32\textwidth]{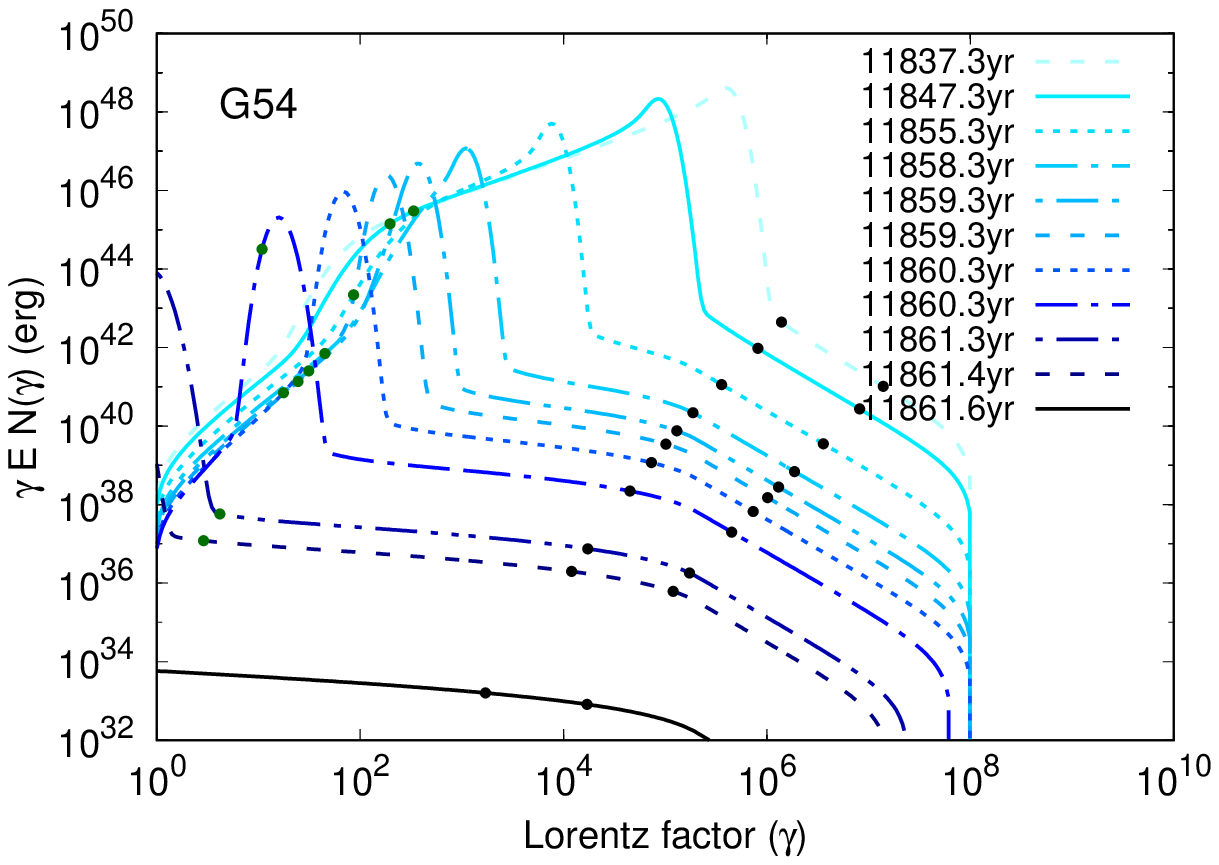}
    \includegraphics[width=0.32\textwidth]{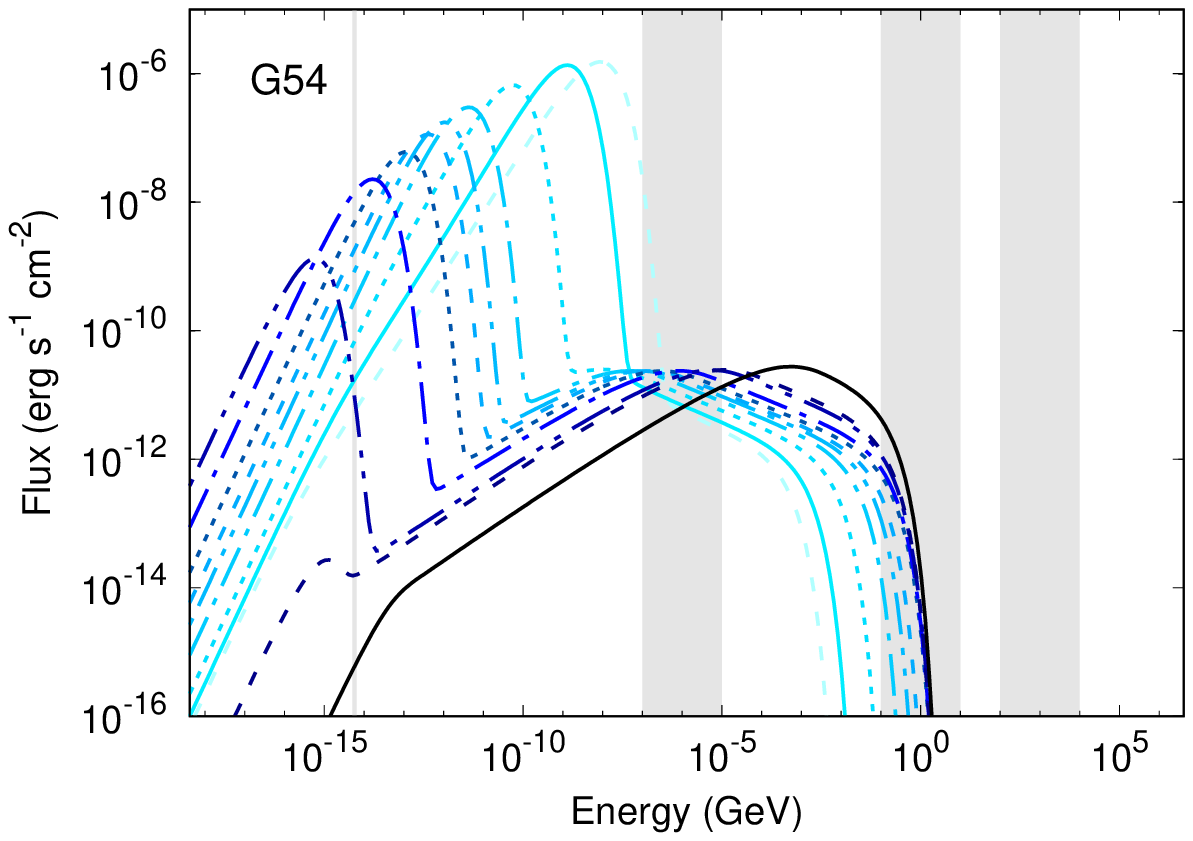}
    \includegraphics[width=0.32\textwidth]{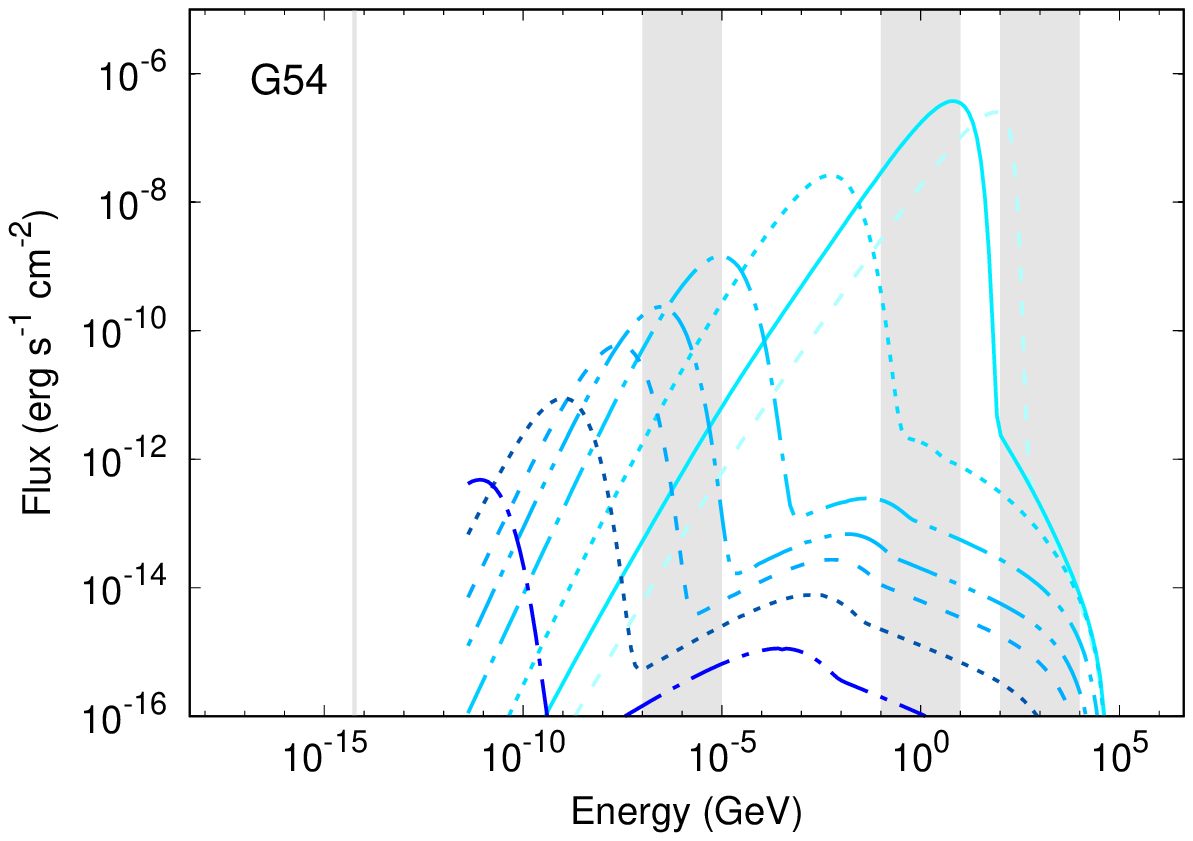}
     \includegraphics[width=0.32\textwidth]{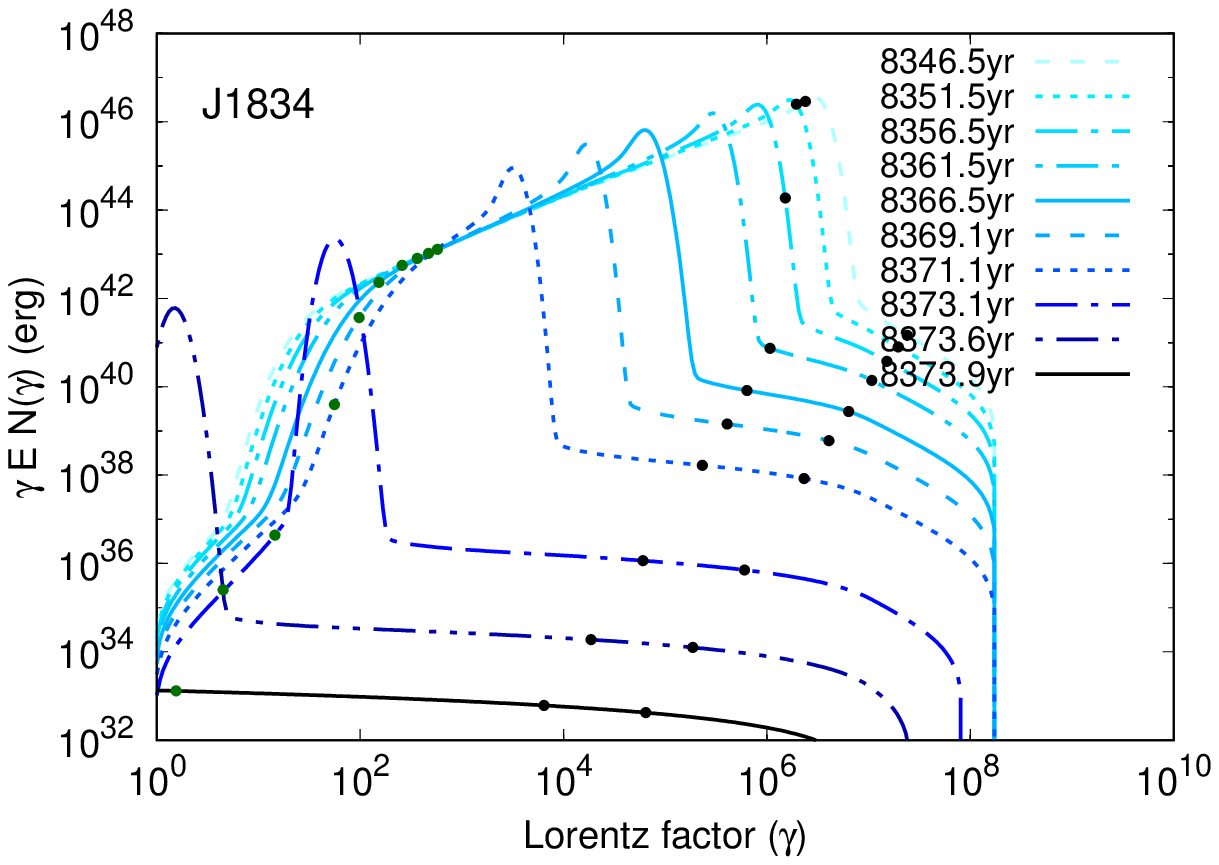}
    \includegraphics[width=0.32\textwidth]{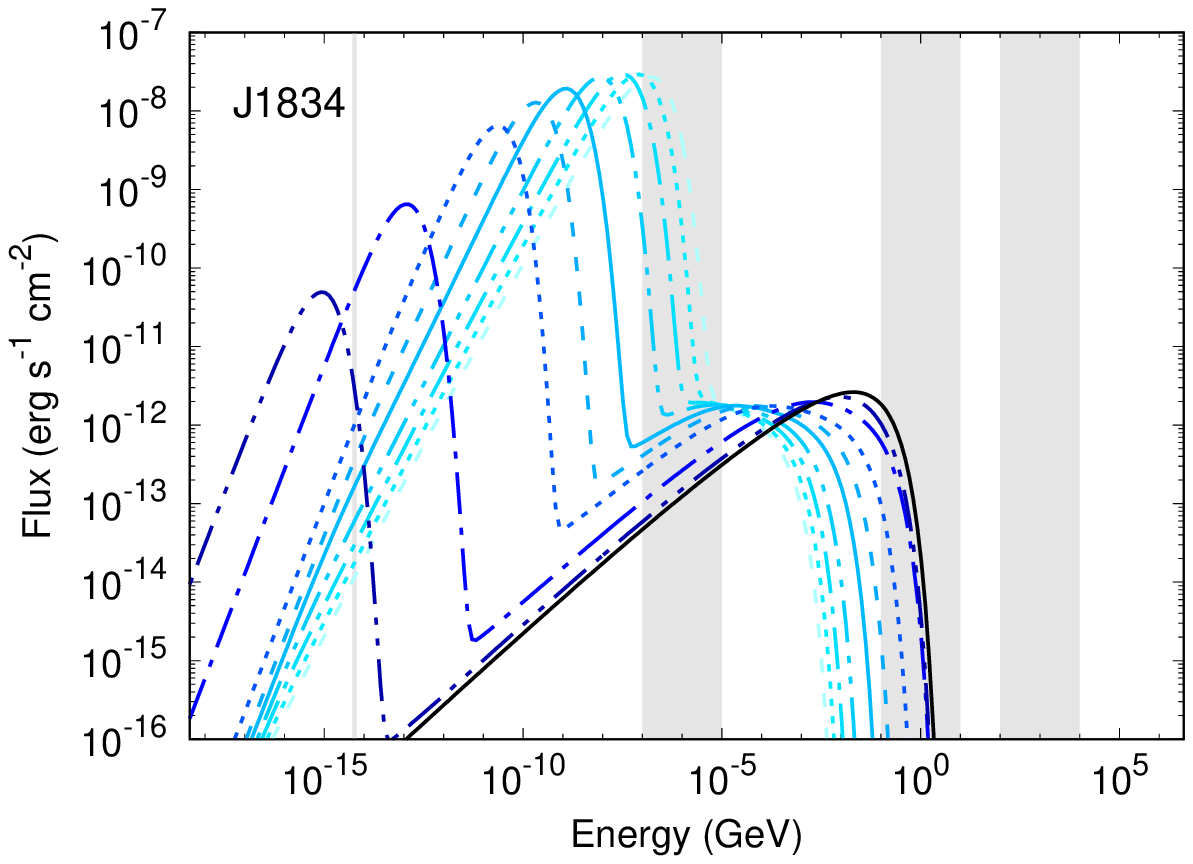}
    \includegraphics[width=0.32\textwidth]{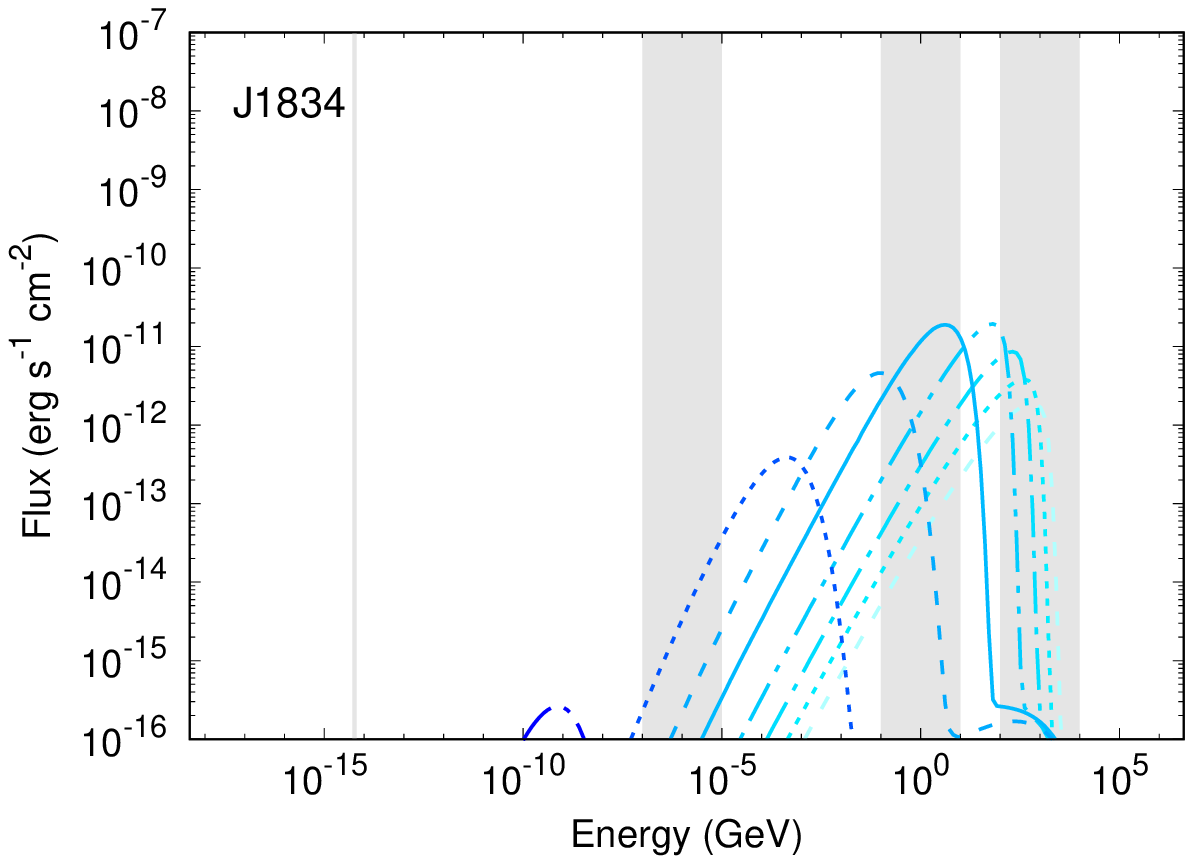}
 \caption{Details of the time evolution of the electron distribution and the synchrotron and self-synchrotron Compton contribution to the photon spectrum around the time of minimum radii and maximum efficiency, between $t_2$ and $t_3$ in the corresponding panels of Figure \ref{time_evolution}. Times are color-coded as described in the left panels.
}
    \label{super-detail}
\end{figure*}

The zoomed panels of
Fig. \ref{super} can actually be considered as a proxy for the evolution of the luminosities  themselves, in arbitrary units. In such a short period of time, the  change 
of the spin-down power is small.
In these zoomed panels, we note the appearance of a second peak in the X-ray efficiency for most of the cases studied (of which those appearing in Crab and G09 are examples).
When such second peak happens, it is closer to the time of the minimal radius. 
Whereas this second peak is however only a local maximum, with the absolute largest X-ray efficiency happening at earlier times,
it may also provide a second -and shorter- superefficiency period in some cases.

In 
the zoomed panels of Fig. \ref{super}, we marked on some exemplary cases (Crab, G09, G54, and J1834) several times of interest between the times of the maximum of the X-ray efficiency and its second local maximum.
At these times, we plotted the electron distribution, the synchrotron and self-synchrotron Compton contribution to the photon spectrum
in Fig. \ref{super-detail}.
This figure shows how  the synchrotron-related processes dominate the shape of the spectrum at both low and high energies
(compare Fig. \ref{super-detail} with the corresponding total SED shown in Fig. \ref{time_evolution}). 
This is particularly obvious when the two peaks in the SED appear clearly distinguished in energy, at the time of maximum efficiency $t_2$.

\begin{figure*}
    \centering
    \includegraphics[width=0.32\textwidth]{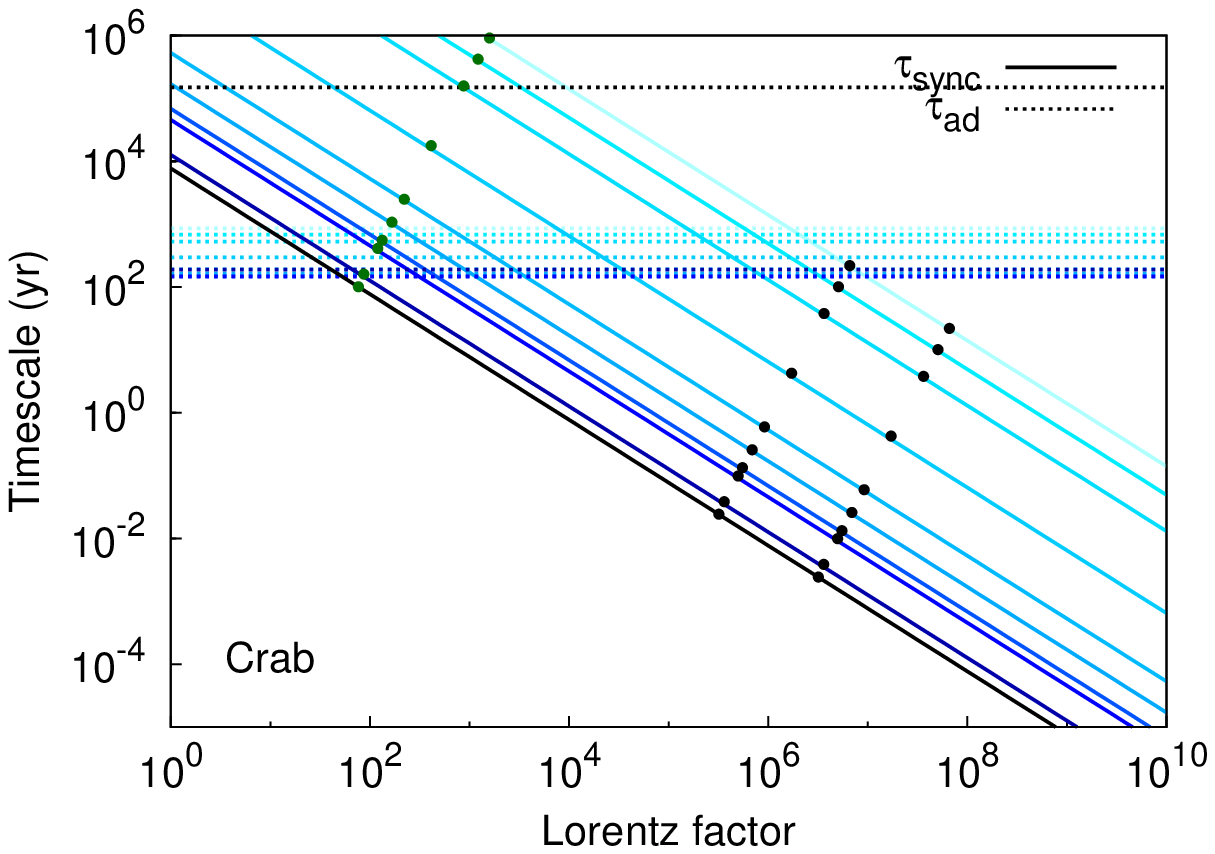}
    \includegraphics[width=0.32\textwidth]{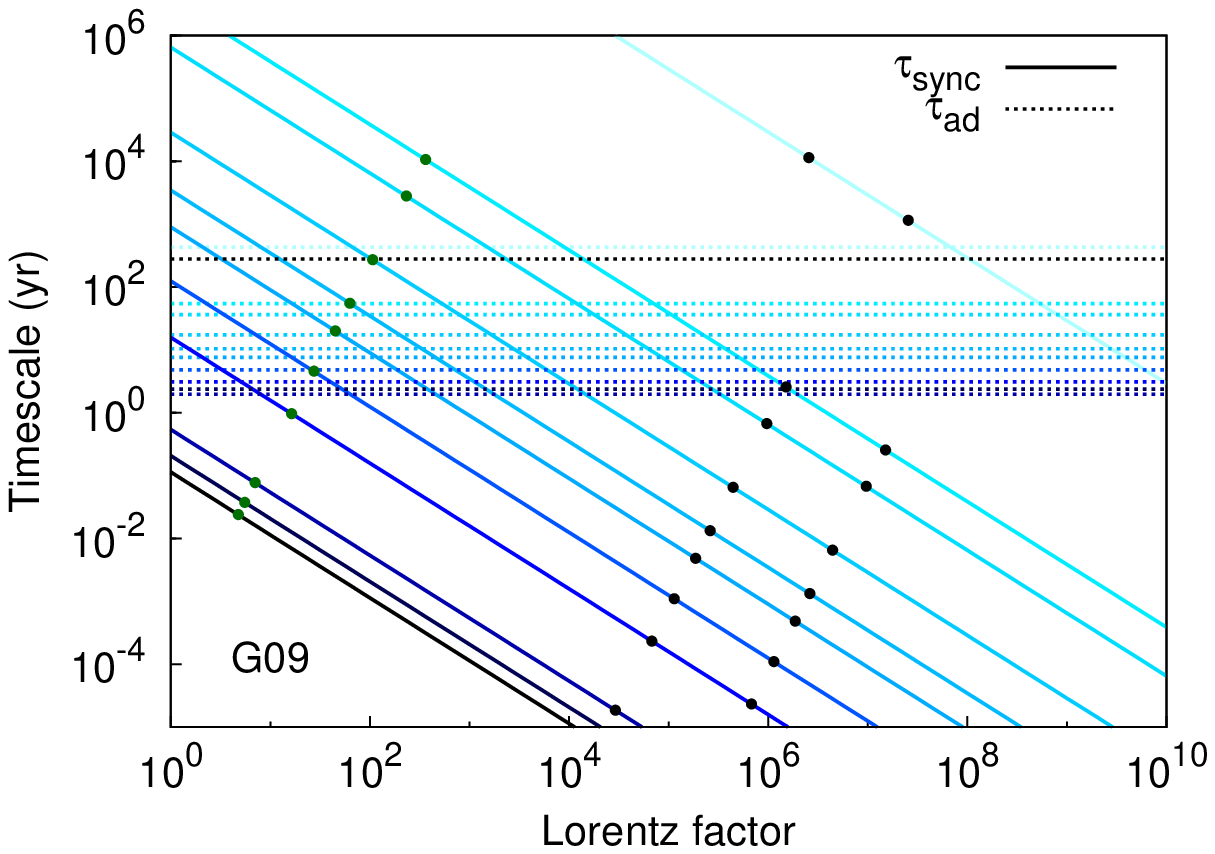}\\
    \includegraphics[width=0.32\textwidth]{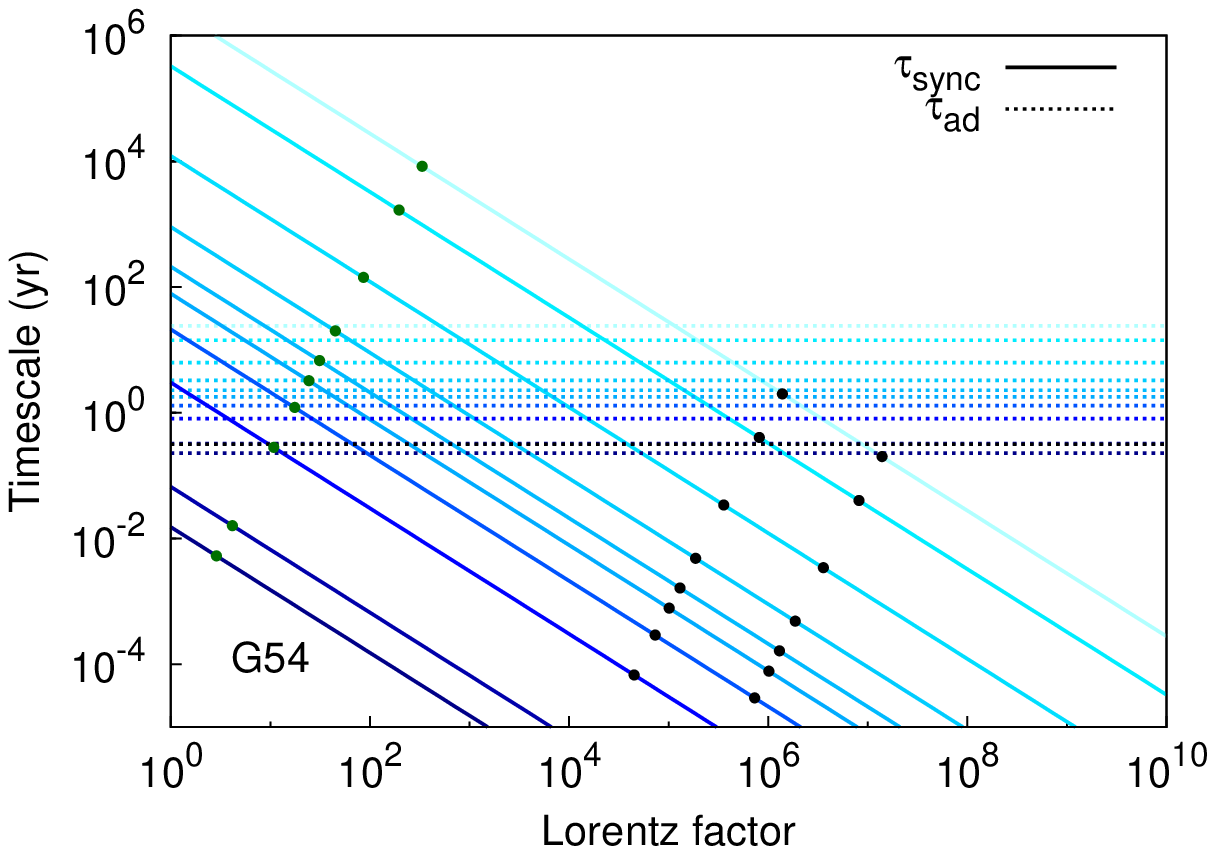}
     \includegraphics[width=0.32\textwidth]{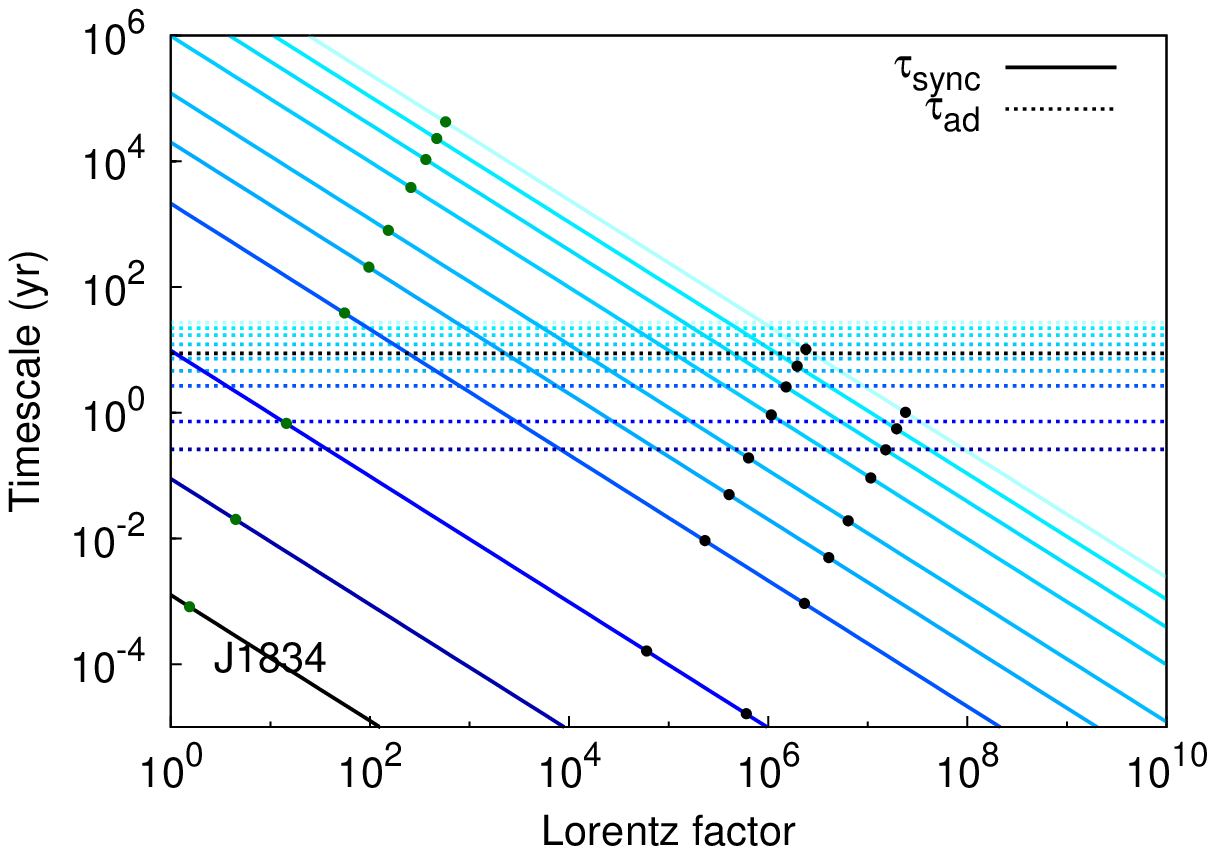}
 \caption{Timescales for the adiabatic heating and the energy losses via synchrotron radiation. The black dots 
 mark the interval of Lorentz factors emitting synchrotron photons with characteristic energies 
in the X-ray band. 
The green dot marks the same for radio (synchrotron photons with characteristic energies at 1.4 GHz).
The curves are color-coded as in Fig. \ref{super-detail}.
}
\vspace{1cm}
    \label{losses}
\end{figure*}

We note that the maximum of the X-ray efficiency does not occur at the minimum of the radius, but at a time in between the start of reverberation and the latter.
This is in remarkable agreement with a result from analytical
considerations earlier done by Bandiera (2014).
This happens in all cases studied, and is a result of the energetic balance: a competence between electron heating by the nebular compression and how fast 
electrons escape or are cooled down via the emission of synchrotron radiation. 
The more compressed is the nebula, electrons are cooled down faster (via synchrotron radiation in a larger magnetic field), 
and a smaller number of electrons are actually available to emit in X-rays. 
The competition between gains by adiabatic heating and losses by synchrotron along the critical time period is shown in Fig. \ref{losses}.
They depict the timescales for energy gains and losses at the same times in which the corresponding SEDs and electron distributions
were shown earlier.
It can be seen that during most of the compression, synchrotron radiation has a shorter timescale than heating for the Lorentz factors of interest, and quickly burns off the electron population in all PWNe.
This is consistent with the SEDs being dominated by synchrotron and self-synchrotron emission, and with the appearance of a second peak in the X-ray efficiency, as further discussed below.

Fig. \ref{super-detail} and \ref{losses} also show the interval of Lorentz factors emitting synchrotron photons with characteristic energies 
($\nu_c=({3}/{4\pi}) \gamma^2 ({eB}/{mc})$)
between 0.1 and 10 keV (noted with black dots), and radio (1.4 GHz, noted with green dots).
The Lorentz factors of interest for the emission of photons at these bands 
change significantly along the time evolution, even in this limited time extent, due to the strong variation in the magnetic field.
%
%To obtain the characteristic energy,$$\nu_c=\frac{3}{4\pi}\gamma^2\frac{eB}{mc}$$, and get $$\gamma^2=2.38\times 10^{-7}\frac{\nu/Hz}{B/G}$$
%
This was also summarily shown in Fig. \ref{time_evolution}, where we showed these intervals at
$t_2=t($Eff$^{max}_{X})$
and
$t_3=t(R_{min})$.
For these Lorentz factors of interest, and along the period shown, the number of electrons uniformly decreases, due to the cooling domination.

At the beginning of phase c the peak of the synchrotron contribution to the SED, shown in the middle panel of Fig. \ref{super-detail}, is close to the X-ray band of interest, affecting the value of efficiency just as a consequence of the band selection.
If, instead, we would be interested in the hard X-ray luminosity above 100 keV, the X-ray synchrotron flux would uniformly increase with time.

In addition of the X-ray luminosity variation via synchrotron, the X-ray flux is also affected by self-synchrotron emission,
see the second and third panels of Fig. \ref{super-detail}.
The latter radiation process dominates the production of the second peak. 
It 
happens at times when the comptonized synchrotron spectrum actually peaks in X-rays instead than in gamma-rays. 
When it does,
the flux in X-rays produced by self synchrotron Compton emission may be one order of magnitude larger than that produced
by synchrotron emission directly.
This emphasizes how important it is to consider the self-synchrotron Compton process along the evolution of all nebulae, even when at later times it may be, usually, completely irrelevant.

Note that when they happen,  these second peaks occur closer to (but still before) the minimum of the radius.
Note too that the GeV (and TeV) maximum efficiency happens always after the X-ray one.
The reason for all this is the same, and is related to the fact that the self-synchrotron emission, which we compute following with the formulae given in the Appendix of Martin et al. (2012), is quadratic in the number of electrons, inversely quadratic in the size of the nebula, and linear in the field. 
The electrons influence is thus larger for self-synchrotron emission, given that they are also accounted in the photon target distribution.
However, the maximum efficiency moves towards later times when compared to the X-ray one since for a longer time the reduction of particles 
is compensated by the increase in the field and the decrease in the radius. 
With reverberation wiping electrons off quickly, once the maximum of the GeV luminosity is attained and starts to decrease, there is no possible compensation to the loss of electrons. 
There is no second peak in GeV or TeV energy bands because at these energy bands there is only one dominant process generating the SED, and the recovery can only happen when a sufficient number of high energy electrons are rebuilt by the pulsar.

%%%%%%%%%%%%%%%%%%%%%%%%%%%%%%%%%%%%%%%%%%%%%%%%%%
\section{Concluding remarks}
%%%%%%%%%%%%%%%%%%%%%%%%%%%%%%%%%%%%%%%%%%%%%%%%%%

Here, we have shown that supereffiency periods in which
the luminosity at a given band from radio to TeV exceeds the pulsar spin-down power, are common. 
They are unavoidably associated with the reverberation process. 
Supereffiency happens because when the PWNe are reverberating, the spin-down power is no longer the energy reservoir.
In these cases, the nebulae are receiving energy from the environment,
and the spin-down power is, a priori, not determinant to judge detectability at any band.

Observing one such superefficient system would be amazing: a bright, small or point-like nebula, with a spatially coincident pulsar many times less energetic. 
The difficulty for observing them is that such systems can be maintained only for a few hundred years. 
For the estimate that follows, let us assume that the superefficiency  period roughly lasts about 300 years in the evolution of young nebulae, of typically $<10000$ years of age (although note that as the G54 case tells, supernova with large ejected
masses or low density environments can produce reverberation beyond this age).
Assuming a pulsar birthrate of 3 century$^{-1}$ \citep{Faucher2006}, 300 PWNe were born within the last $10000$ years,
and from these, we are interested in a period equivalent to --at most-- 3\% of their evolution.
Taking into account the correspondingly shorter percentages for pulsars born at different centuries, 
we have a probability of $\sim 1\%$ of finding one these pulsars in the right period of their evolution.
Thus we expect at most 3 PWNe in a superefficient stage in the Galaxy today.
This should be taken rather as an upper limit, because it assumes it is equally probable to have reverberation at any time within 
the first 10000 years of a pulsar (thus neglecting that there is no reveberation in their free-expansion phases).
In a future work, we shall focus on observational strategies for finding superefficient or highly efficient PWNe.

Note that our model assumes no morphological shape for the PWN; they are described with a time-varying radius.
If the compression is asymmetric or turbulence develops, superefficiency could be less effective,
detaining the reduction in the PWN size and the increment in the field perhaps before our results indicate.
This might affect less energetic nebulae in particular, such as Kes 75 or J1834, 
being likely unimportant for others such as G09, G21 or Crab.
Magneto-hydrodynamical simulations will verify on this issue.
In any case, $R_{min}$ is many orders of magnitude larger than the pulsar's radius, or even the pulsar's magnetosphere (typically at least 6 orders of magnitude larger than the size of a young pulsar's light cylinder), and 
thus the inner workings of the pulsed emission via synchro-curvature radiation \citep{Torres2018}, is not expected to be significantly affected even in the most severe of the compressions.

%%%%%%%%%%%%%%%%%%%%%%%%%%%%%%%%%%%%%%%%%%%%%%%%%%
\acknowledgements
%%%%%%%%%%%%%%%%%%%%%%%%%%%%%%%%%%%%%%%%%%%%%%%%%%

This research was supported by the grants  AYA2015-71042-P, SGR2017-1383, iLink 2017-1238, and the National Natural Science Foundation of
China via NSFC-11473027, NSFC-11503078, NSFC-11133002, NSFC-11103020, NSFC-11673013, XTP project XDA 04060604 and the Strategic Priority Research Program ``The Emergence of Cosmological Structures" of the Chinese Academy of Sciences, Grant No. XDB09000000.

\begin{table*}
\renewcommand{\arraystretch}{0.7}%
\scriptsize
    \centering

\caption{Physical parameters used by, and resulting from the fits. 
}

\begin{tabular}{lllllllll}\hline\hline

PWN  & &  {\red {Crab Nebula    }} &  {\red {G0.9+0.1 }}  & {\red { G21.5-0.9 }}  &  {\red {G54.1+0.3 }}    &  {\red {G29.3-0.3   }}    &  {\red {J1834.9-0846  }}    & \\
referred to as  & &  {\red {Crab    }} &  {\red {G09    }}  & {\red { G21    }}  &  {\red {G54  }}    &  {\red {Kes 75   }}    &  {\red {J1834   }}    & \\ \hline

\hline
 {\blue {Measured or assumed }}\\
\hline

$t_{age}$ & [years]     & 960   & 2000    & 870    & 1700    & 700    & 7970    &\\

${P}(t_{age})$ & [ms] &33.40                &52.2                &61.86                &136                &324                &2480                &\\

$\dot{P}(t_{age})$ & [s s$^{-1}$] &$4.2\times10^{-13}$&$1.5\times10^{-13}$&$2.0\times10^{-13}$&$7.5\times10^{-13}$&$7.1\times10^{-12}$&$7.96\times10^{-12}$&\\

 $n$            &   \ldots    & 2.509               & 3                   & 3                   & 3                   & 2.16                & 2.2                 &\\

 $D$ & [kpc]           & 2.0                 & 8.5                 & 4.7                 & 6                   & 6                   & 4                   &\\

 \hline
 {\blue { Derived }}\\
 \hline

 $\tau_{0}$ & [years]     &  758                & 3305                & 3985                & 1171                & 547                 & 280                 &\\

 $B_s$ & [G] & $4.68\times10^{12} $        & $2.91\times10^{12} $    & $3.58\times10^{12} $      & $1.01\times10^{13} $    & $8.19\times10^{13} $     & $1.43\times10^{14} $    &\\

 $L_{sd}(t_{age})$ & [erg s$^{-1}$] &$4.5\times10^{38}$   &$4.3\times10^{37}$   &$3.4\times10^{37}$   &$1.2\times10^{37}$   &$8.2\times10^{36}$   &$2.1\times10^{34}$   &\\

 $L_{0}$ & [erg s$^{-1}$] & $3.0\times10^{39}$ & $1.1\times10^{38}$ & $5.0\times10^{37}$ & $7.2\times10^{37}$ & $7.7\times10^{37}$ & $1.74\times10^{38}$ &\\

 \hline
 {\blue {Fitted parameters}}\\
 \hline

 $M_{ej}$ & [M$_{\odot}$] &  9.0               & 11                 & 8                 & 20                 & 6                 & 11.3               &\\

  $T_{fir}$ & [K]       & 70                  & 30                  & 35                  & 20                  & 25                  & 25                  &\\
 $\omega_{fir}$ & [eV  cm$^{-3}$] & 0.1             & 2.5         & 1.4         & 2.0         & 2.5         & 0.5         &\\
 $T_{nir}$ & [K]       & 5000                & 3000                & 3500                & 3000                & 5000                & 3000                &\\
 $\omega_{nir} $ & [eV cm$^{-3}$] &  0.3             & 25          & 5.0          & 1.1          & 1.4          & 1.0         &\\
 $\gamma_b$   & \ldots      &  $9\times10^5$   & $1.0\times10^5$      & $1.0\times10^5$       & $5.0\times10^5$       & $2.0\times10^5$       & $1.0\times10^7$     &\\
 $ \alpha_l$    & \ldots    & 1.5                 & 1.4                 & 1.0                 & 1.2                 & 1.4                 & 1.0                 &\\
 $\alpha_h$     & \ldots    & 2.54            & 2.7                 & 2.5                 & 2.8                 & 2.3                 & 2.1                 &\\
 $\epsilon$     & \ldots    &  0.27            & 0.2                 & 0.2                 & 0.3                 & 0.2                 & 0.6                 &\\
 $\eta$          & \ldots   &  0.02 & 0.01                & 0.04                & 0.005                & 0.008                & 0.045               &\\

 \hline

 {\blue {Resulting features}}\\

\hline

Rev. Timescale  &  [years]    & 3201.7    & 1703.9    & 1028.86    & 2242.61    & 674.679    & 1037.906    &\\

$t_1=t(R_{max})$ & [years] & 3640    & 5341    & 4549    & 9619    & 4007    & 7336    & \\

%\hdashline

$r_{min}$  & [pc] &  $9.8\times10^{-2}$   & $3.2\times10^{-3}$    & $7.3\times10^{-3}$    & $1.3\times10^{-4}$    & $7.5\times10^{-5}$    & $5.5\times10^{-4}$    &\\

$B_{max} = B_{t_3}$  &  [G]   & $5.6\times10^{-2}$    & $14.7$    &  $4.7$   & $2.0\times10^{3}$    & $3.9\times10^{3}$    & $1.4\times10^{2}$    &\\

%\hdashline

Eff.$^{max}_{\rm r}$ & \ldots & 4.1 & 41.9    & 7.2    & 106.9    & 905.5    & 14.9     &\\
Dos$_{\rm r}$&  [years]    &  150   & 12.2    & 4.0    &  4.0   & 3.9    & 1.0    &\\
$t($Eff.$^{max}_{\rm r})$ & [years]    &  6834.2    & 7041.8    & 5575.8    &    11860.8 & 4682.5    & 8373.5    &\\

%\hdashline

Eff.$^{max}_{X}$ & \ldots   & 2.9    & 54.2    & 36.8    & 5464    & 23523    & 5444    &\\
Dos$_{X}$ & [years]   & 800    & 327    & 237    & 342    & 129    & 316    &\\
$t_2=t($Eff$^{max}_{X})$ & [years]    &  5700    & 6900    & 5450    & 11807    & 4668    & 8337    &\\

%\hdashline

Eff.$^{max}_{\rm GeV}$ &  \ldots  & 11.7    & 1180    & 100.8    & 6541    & 22681    & 5.5    &\\
Dos$_{\rm GeV}$ & [years]   & 439    & 127    & 67    & 58    & 24    & 8    &\\
$t($Eff$^{max}_{\rm GeV})$ &  [years]   &  6500    & 7007    & 5547    & 11847    & 4680    & 8366    &\\

%\hdashline

Eff.$^{max}_{\rm TeV}$   & \ldots & 0.9    & 72.4    & 6.6    & 1193    & 0.01    & 1.00    &\\
Dos$_{\rm TeV}$&  [years]    &  \ldots   & 92    & 37    & 147    & 29    & 5    &\\
$t($Eff.$^{max}_{\rm TeV})$ & [years]    &  6250    & 6960    & 5510    & 11832    & 4680    & 8356    &\\

%\hdashline

$t_3=t(R_{min})$  & [years]  & 6841.7    & 7044.9    & 5577.9    & 11861.6    & 4682.7    & 8373.9    &\\

$t_4=t(@Sedov)$ & [years] & 8011.4    & 8002.9    & 5993.1    & 13025.7    & 5007.6    & 8988.8    & \\

%\hdashline

$B_{t_4}$ & [G]   & $2.2\times10^{-4}$    & $2.2\times10^{-4}$    & $2.7\times10^{-4}$    &  $1.7\times10^{-4}$   & $3.0\times10^{-4}$    & $2.1\times10^{-4}$    &\\

$B_{today}$ & [G]   & $1.1\times10^{-4}$    & $1.4\times10^{-5}$    &  $8.5\times10^{-5}$   & $1.2\times10^{-5}$    & $2.5\times10^{-5}$    & $4.9\times10^{-6}$    &\\

 \hline

\end{tabular}

\label{table1}
\end{table*}

%\bibliography{pwn-bib}

\end{document}